\documentclass[aip,adv,reprint,superscriptaddress,amsmath,amssymb,floatfix]{revtex4-2}
\usepackage[final]{graphicx}
\usepackage{amssymb}
\usepackage{amsmath}
\usepackage{graphicx}
\usepackage[export]{adjustbox}
\usepackage{float}
\usepackage{wrapfig}
\usepackage{xcolor}
\usepackage{gensymb}
\usepackage{hyperref}
\hypersetup{
     colorlinks   = true,
     citecolor    = blue,
     linkcolor   = blue
}
\usepackage{comment}
\usepackage[utf8]{inputenc}
\usepackage[T1]{fontenc}
\usepackage{mathptmx}
\usepackage{etoolbox}

\makeatletter
\def\@email#1#2{%
 \endgroup
 \patchcmd{\titleblock@produce}
  {\frontmatter@RRAPformat}
  {\frontmatter@RRAPformat{\produce@RRAP{*#1\href{mailto:#2}{#2}}}\frontmatter@RRAPformat}
  {}{}
}%
\makeatother

\emergencystretch=\maxdimen
\begin{document}

\title{Data driven screening of CaCu$_5$-type magnetic structures}
\author{Nabaraj Pokhrel}
\affiliation{Materials Science and Technology Division, Oak Ridge National Laboratory, Oak Ridge, TN, 37831, USA}
\author{Sheila Whitman}
\affiliation{Department of Applied Mathematics, University of Arizona, Arizona, 85721, USA}
\author{David S. Parker}
\email{pokhreln@ornl.gov}
\affiliation{Materials Science and Technology Division, Oak Ridge National Laboratory, Oak Ridge, TN, 37831, USA}

\date{\today}

\begin{abstract}
We develop a machine learning assisted workflow for the rapid screening of permanent magnet candidates within the CaCu$_5$ structure family. We generated more than 60,000 structures through substitutions. These compositions were screened using a combination of pretrained materials models, Materials Project data, and a machine-learning model for magnetization prediction. Four criteria, namely, metallicity, negative formation energy, energy above the convex hull $\leq 50$~meV/atom, and predicted magnetization above the adopted $\geq 0.75$~T threshold reduced the initial chemical space to 439 candidates for first-principles calculations. Density functional theory calculations revealed that 159 compounds exhibit positive axial magnetic anisotropy energy, favoring [001] over the tested in-plane directions. The resulting dataset reveals chemical trends in magnetic anisotropy across the CaCu$_5$ compositional space and provides a set of candidates for further investigation. More broadly, this work demonstrates a practical strategy for combining existing machine learning models with first-principles calculations to systematically reduce large chemical spaces to tractable sets for computationally intensive materials screening.

\end{abstract}

\maketitle

\section{Introduction}

Data driven approaches, particularly machine learning (ML), are increasingly being used for the discovery of new materials and the optimization of existing ones. However, their application remains challenging in several areas and requires further demonstrations of their practical utility, including in the design of rare-earth (RE) permanent magnets. The vast and complex compositional space of these materials, combined with the limited availability of high quality experimental data, presents significant challenges for developing reliable ML based approaches. Widely followed conventional trial-and-error approaches are time consuming and computationally demanding, whereas ML assisted methods may enable rapid screening of millions of candidate compositions and crystal structures. Recently, ML guided first-principles studies have gained significant traction in the field of magnetism~\cite{Iwasaki2021, Pant2023, Katsikas_2021, Xia_2022}. Similar strategies have been successfully applied to predict key properties of RE based magnets~\cite{PhysRevApplied.14.034024, PhysRevApplied.22.024046} and to explore RE free permanent magnet materials~\cite{10.1063/5.0208761}. As ML models typically require large datasets for effective training, one practical approach to data generation involves systematic atomic substitutions to create synthetic datasets~\cite{Pant2023}. This strategy accelerates the identification of promising candidates while minimizing the need for extensive experimental synthesis. When integrated with high-throughput experimental workflows, such ML driven approaches can form the basis of a closed loop materials discovery framework that continuously refines and improves predictive accuracy.

Permanent magnets based on the CaCu$_5$-type crystal structure have been extensively studied through both first-principles calculations and experimental investigations~\cite{10.1063/1.1709630, PhysRevMaterials.5.124405, Samolyuk_2024}. SmCo$_5$ is a well known example of this structure family, exhibiting strong uniaxial magnetic anisotropy with a magnetocrystalline anisotropy energy ($K_1$) of approximately 17 MJ/m$^3$ and a Curie temperature ($T_C$) exceeding 1000 K~\cite{buschow1969preparation, ma17040808}. In SmCo$_5$, the large magnetic anisotropy primarily originates from the contribution of Sm 4$f$ electrons. Notably, CaCu$_5$-type compounds that do not rely on 4$f$ electrons, such as YCo$_5$ and LaCo$_5$, also display $K_1$ values comparable to or exceeding those of state-of-the-art NdFeB magnets. In these materials, the origin of the magnetocrystalline anisotropy is more complex. For example, in YCo$_5$, the uniaxial anisotropy arises from spin-orbit coupling mediated hybridization between the Co $d_{x^2-y^2}$ and $d_{xy}$ orbitals~\cite{Samolyuk_2024}. These characteristics make the CaCu$_5$-type structure family particularly attractive for further exploration using ML approaches, which can help identifying the key descriptors and design principles needed to optimize magnetic performance.

\begin{figure}[tbp]
\centering
    \includegraphics[width=\columnwidth]{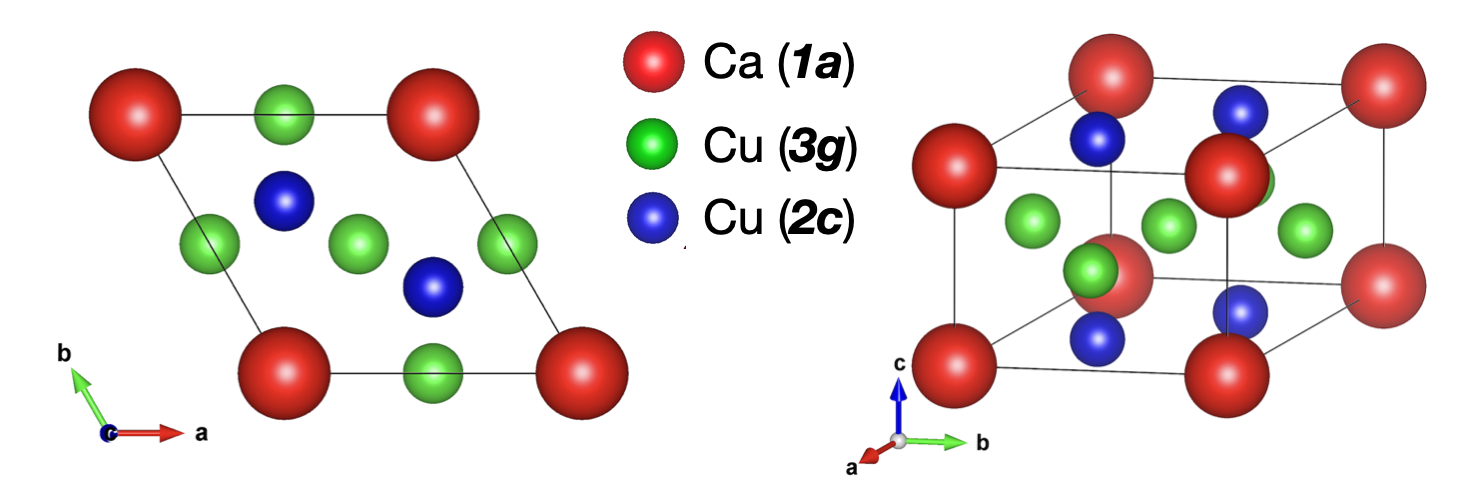}
    \caption{Crystal structure of CaCu$_5$ in its hexagonal phase with $P6/mmm$ symmetry with color coded Wyckoff positions.\label{fig:CaCu5_crystal_1}}
    
\end{figure}

The CaCu$_5$ structure consists of alternating layers of Ca and Cu, as illustrated in Figure~\ref{fig:CaCu5_crystal_1}, and crystallizes in the hexagonal space group $P6/mmm$. This structural motif extends to a broader family of compounds through substitution of Ca with RE elements such as Y, La, and Sm, and Cu with other transition metals, particularly Co. However, relatively few CaCu$_5$-type compounds have been experimentally synthesized to date~\cite{10.1063/1.106737, ONYSZCZAK2019192, YCo5_expt_1966}, resulting in a limited dataset for data driven materials design. To overcome this limitation for this work, we use a systematic atomic substitution strategy to generate a large and diverse set of hypothetical CaCu$_5$-type structures. They contain three symmetry distinct Wyckoff sites, $1a$, $3g$, and $2c$, which can be independently substituted with elements across the periodic table. This approach enables the systematic generation of thousands of chemically diverse structures, providing a sufficiently large candidate dataset for systematic materials screening and discovery.

Here, we develop and assess a reusable workflow for rapidly screening this chemical space. The goal is to evaluate how existing ML based tools can be integrated with first-principles density functional theory (DFT) calculations to develop flexible workflows that can be readily applied or adapted to the needs of different projects. Rather than replacing DFT or directly establishing experimental synthesizability, these workflows are intended to accelerate computational screening, identify promising candidates, and support both computational and experimental efforts. We combine pretrained graph-network models, database derived phase-diagram estimates, and a magnetization regressor, followed by DFT calculations. The resulting dataset identifies promising site chemistry while also revealing the limitations of the screening approach and areas requiring further validation.

\section{Methods\label{sec:method}}

\subsection{Candidate generation and ML screening}

Our atomic substitution strategy generated 84,564 unique CaCu$_5$-type candidate identifiers after duplicate removal. Of these, 62,017 compositions have all four screening quantities available in the archived dataset. The remaining candidates were not compatible with one or more stages of the screening workflow, for example, because their compositions were not supported by the M3GNet model. All candidate structures were initialized in the $P6/mmm$ CaCu$_5$ prototype using the same six atom unit cell. Details regarding the starting geometry, supported element restrictions, and descriptor details are provided in Sections~I, I.1, and II, respectively, of the Supplemental Material~\cite{supp}. 

\begin{figure*}[t]
\centering
    \includegraphics[width=0.8\textwidth]{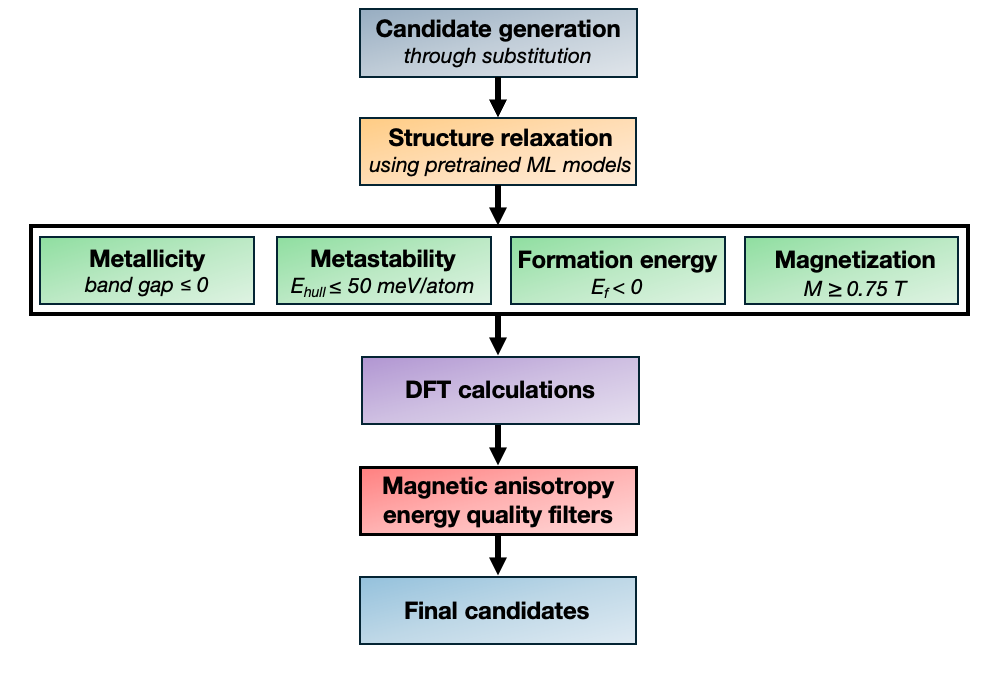}
    \caption{Machine learning-DFT workflow used for this project.\label{ML_DFT_workflow.png}}
\end{figure*}

The screening workflow is shown in Figure~\ref{ML_DFT_workflow.png}. M3GNet-MP-2021.2.8-DIRECT-PES was used for structural relaxation and to obtain candidate energies~\cite{M3GNET,qi2024robust}. MEGNet-MP-2019.4.1-BandGap-mfi was used to predict band gaps, with candidates having predicted band gaps $\leq0$ classified as metallic~\cite{Chen2019,chen2021learning}. Screening level formation energies and energies above the convex hull were estimated by combining the ML derived candidate energies with competing phase entries and elemental reference energies from the Materials Project~\cite{materials_project}. Because the candidate and competing phase energies may not share identical data provenance or correction schemes, these quantities were used only as screening level stability metrics rather than definitive thermodynamic predictions.

Magnetization was predicted using a random-forest regressor trained on Materials Project data. The input features included lattice parameters, number of atoms, elemental fractions, and composition-based descriptors from Matminer~\cite{ward2018matminer}. The final model used 300 trees, a maximum depth of 30, a minimum split size of 10, a minimum leaf size of 4, and $\sqrt{N_{\rm features}}$ feature sampling (see Supplemental Material Section~II and Table~S1).

Four screening criteria were evaluated independently on the same candidate pool and then combined by intersection. Of the 62,017 jointly evaluable candidates, 52,864 were classified as metallic, 5,984 had $\Delta E_{\rm hull}\leq50$~meV/atom, 34,119 had negative predicted formation energies, and 3,102 had predicted magnetization $\geq0.75$~T. Applying all four criteria simultaneously yielded 439 candidates for first-principles calculations (see Supplemental Material Section~I.2).

\subsection{Magnetization model evaluation}

Several regression models were evaluated for predicting magnetization, including linear regression, ridge regression, support vector regression (SVR), and random forest. Their performance was compared using fivefold cross validation, with the fold averaged results from this initial, untuned comparison summarized in Table~\ref{tab:model-performance}. The random-forest model gave the lowest mean absolute error (MAE) and root mean square error (RMSE), and the highest $R^2$ among the four models and was therefore selected for further optimization.

The random forest hyperparameters were subsequently optimized using randomized search cross validation. The final model was evaluated using an 80:20 train:test split containing 123,893 training and 30,974 test structures. Because hyperparameter selection was performed using the same dataset rather than within a nested cross validation framework, the resulting test set errors should be interpreted as internal model selection metrics rather than as a prospective external benchmark.

\begin{table}[t]\caption{Fivefold cross validation means for the initial magnetization regression comparison. Mean absolute error (MAE) and root mean square error (RMSE) are on the historical training target scale, nominally in tesla.}\label{tab:model-performance}\centering
\begin{tabular}{lccc}\hline\hline
Model & MAE (T) & RMSE (T) & $R^2$ \\\hline
Linear regression & 0.1249 & 0.1986 & 0.6875 \\
Ridge regression & 0.1249 & 0.1986 & 0.6875 \\
SVR (RBF kernel) & 0.0897 & 0.1575 & 0.8036 \\
Random forest & 0.0597 & 0.1405 & 0.8437 \\\hline\hline
\end{tabular}\end{table}

\subsection{First-principles calculations}

Screening level first-principles DFT calculations were performed using the Vienna Ab initio Simulation Package (VASP)~\cite{kresse95} within the Perdew-Burke-Ernzerhof (PBE) generalized gradient approximation~\cite{PBE_functional} and the projector augmented wave (PAW) method~\cite{PAW_method}. The requested electronic-energy and ionic-force convergence thresholds were $10^{-7}$~eV and 10~meV/\AA, respectively. The screening calculations used a 500~eV plane-wave cutoff and the six atom starting unit cell with a $\Gamma$-centered $4\times4\times4$ $k$-point mesh. On-site Hubbard corrections were applied to the RE $4f$ and transition metal $d$ states following the parameters reported by Topsakal and Wentzcovitch~\cite{topsakal2014accurate} and the Materials Project conventions~\cite{materials_project} (see Supplemental Material Section~III.2.1 and Table~S4).

Both parallel and antiparallel initial alignments between the RE and transition metal magnetic sublattices were considered during structural relaxation. Following relaxation, non-self-consistent spin-orbit coupling (SOC) calculations using fixed charge densities were performed with the magnetization oriented along the [001], [100], and [010] directions. The magnetocrystalline anisotropy energy was defined as

\begin{equation}
K_1 =\frac{E_{[100]}-E_{[001]}}{V},
\end{equation}

\noindent
where $V$ is the volume of the relaxed unit cell. With this convention, $K_1 > 0$ favors [001] over [100]. We use $K_1$ as the screening level axial anisotropy energy density, identifying it with the first-order anisotropy constant and assigning an easy axis assumes negligible higher order terms. The [001] magnetic moments used in the statistical analysis are the magnitudes of the combined atom projected spin and orbital contributions, $|m_{\rm tot}^{\rm proj}|=|m_s^{\rm proj}+m_l^{\rm proj}|$. As pragmatic screening level quality checks, structures were excluded when the in-plane energy difference exceeded $0.10$~MJ/m$^3$, that is,
$|E_{[100]}-E_{[010]}|/V > 0.10$~MJ/m$^3$, or when $|K_1|>30$~MJ/m$^3$. These filters do not establish full $k$-mesh convergence, a magnetic ground state, or thermodynamic stability, and the magnitude cutoff can also exclude physically large anisotropies (see Supplemental Material Section~III.2.2 and Eqs.~S3--S4).

\section{Results\label{sec:results}}

\subsection{Screening model diagnostics}

We first evaluated the screening models on known CaCu$_5$-type compounds from the Materials Project as a prototype specific diagnostic. Of the 204 structures identified, 25 contained elements not supported by the pretrained models, leaving 179 structures for evaluation. Among these, 97.8\% of the known metallic compounds were classified as metallic by MEGNet using the zero gap threshold. Because this set contains no known nonmetallic compounds, this result serves as a positive control check instead of a general assessment of metal or nonmetal classification performance.

We also evaluated the magnetization model on the known CaCu$_5$-type structures. The model yielded an MAE of 0.0399~T, an RMSE of 0.1148~T, and $R^2=0.9409$. However, this diagnostic is not fully independent because the 204 known structures were not excluded before constructing the general train-test split and may therefore overlap with the training data. For comparison, on the random 20\% test set used for the final model, the random forest achieved an MAE of 0.0664~T, MSE of 0.0200~T$^2$, RMSE of 0.1414~T, and $R^2=0.8361$. These results demonstrate useful predictive performance but also highlight the limitations of random structure level splitting and motivate more stringent composition or prototype grouped validation for future screening studies.

\begin{figure*}[t]\centering
\includegraphics[width=0.85\textwidth]{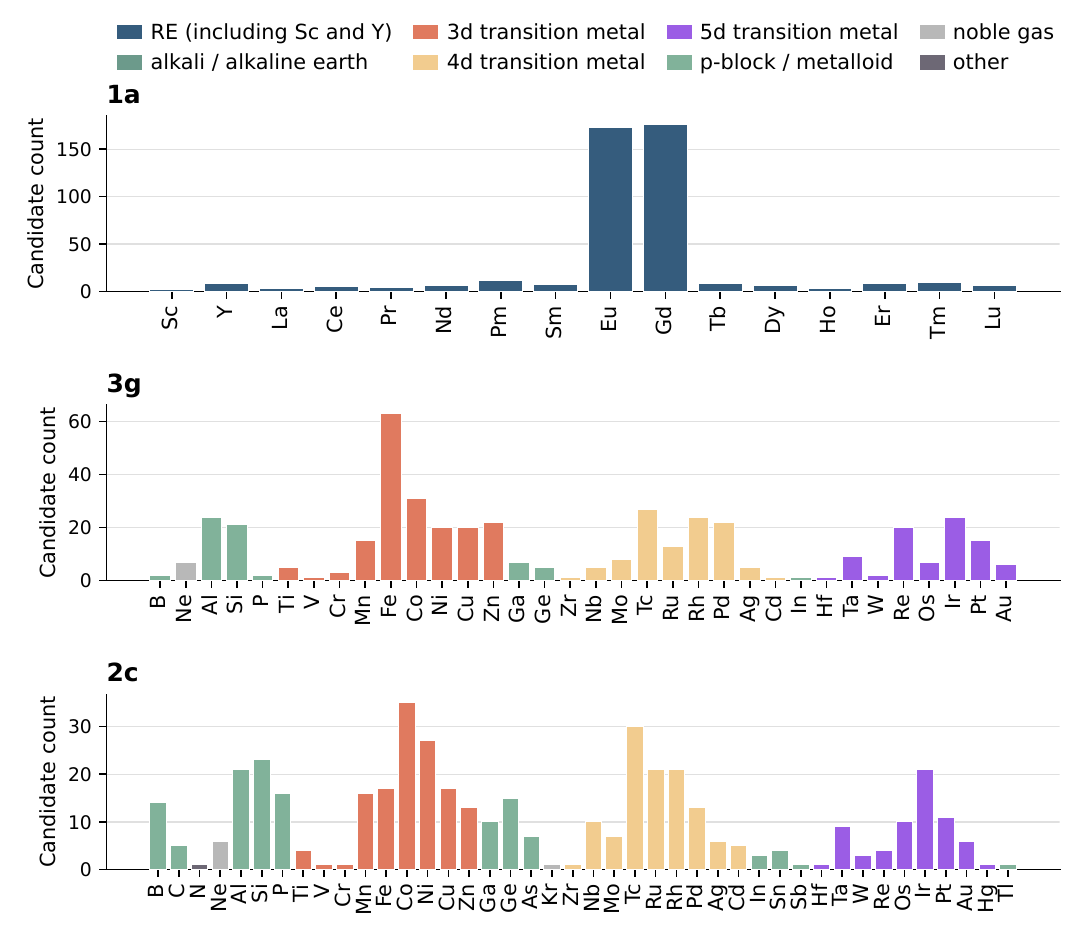}
\caption{Element frequencies at the $1a$, $3g$, and $2c$ sites in the 439 screened CaCu$_5$-type structures. Li, Mg, and Sr are omitted only from the $1a$ panel; their candidates remain included in the other panels and in the screened dataset. Colors indicate broad chemical groups.}\label{fig:frequencies}
\end{figure*}

\subsection{Screened chemical space}

Because RE elements can provide substantial contributions to the magnetocrystalline anisotropy through single ion anisotropy, we focus here primarily on candidates containing RE elements at the $1a$ site. The four screening criteria reduced the initial chemical space to 439 unique compositions selected for subsequent DFT calculations. Figure~\ref{fig:frequencies} shows the elemental distributions across the three crystallographic sites. Gd and Eu are the most frequently occurring elements at the $1a$ site, while Fe and Rh are the most common elements at the $3g$ and $2c$ sites, respectively. Five Ce containing and seven Sm containing compositions passed all four screening criteria, with their compositions provided in Supplemental Material Section~III.3 and Table~S6. As a positive-control check of the screening approach, the known compounds CeCo$_5$, LaCo$_5$, SmCo$_5$, and YCo$_5$ also passed all four ML based criteria, indicating that the data driven screening funnel can recover established CaCu$_5$ type permanent magnet compositions.

\begin{figure*}[t]\centering
\includegraphics[width=0.95\textwidth]{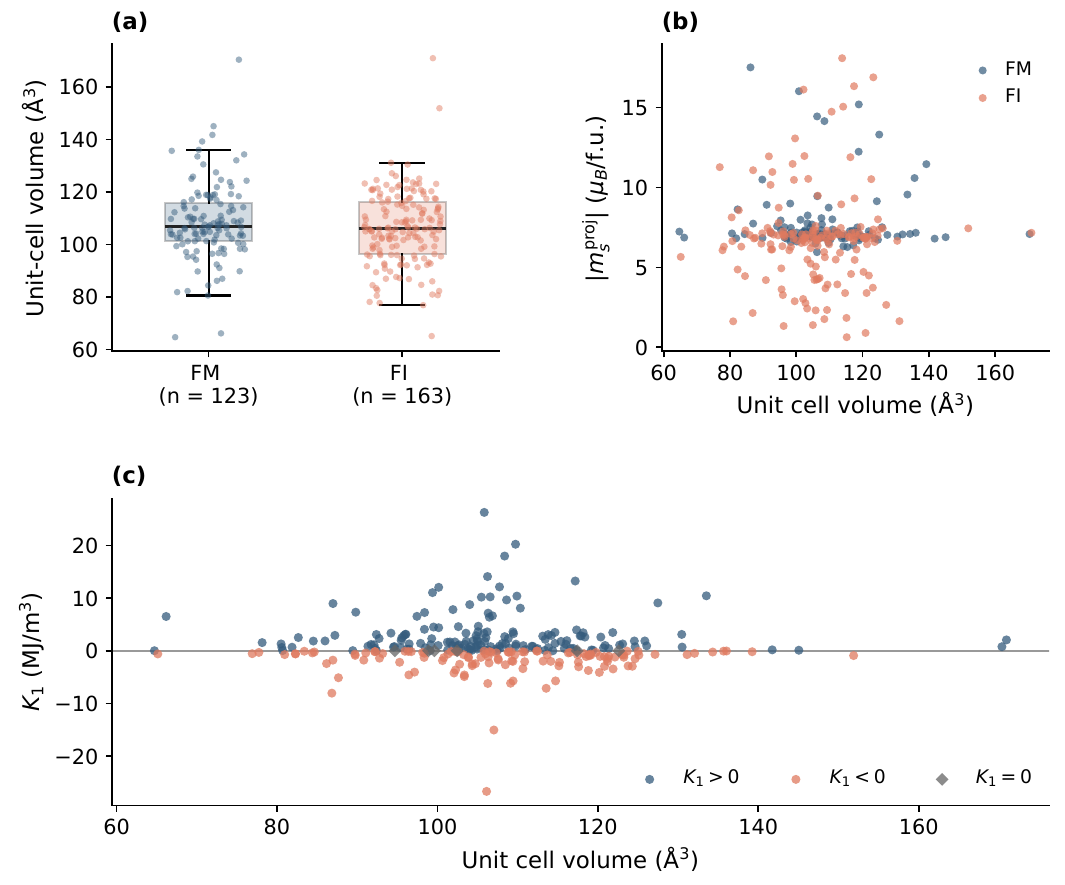}
\caption{Unit cell volume versus (a) lower energy magnetic initialization branch, (b) magnitude of the [001] combined projected total magnetic moment, and (c) axial anisotropy energy $K_1$ for the 286 retained compounds, with 285 complete moments in (b). FM and FI denote parallel and antiparallel alignments of RE and transition metal spins, respectively. Points in (a) show individual compounds where the boxes indicate the interquartile range and median, with whiskers extending to 1.5 times the interquartile range. In (c), colors distinguish positive, negative, and numerically zero $K_1$.}\label{fig:correlations}
\end{figure*}

After applying the anisotropy quality control criteria, 286 compounds remained for analysis (see Supplemental Material Section~III.4 and Table~S7). Of these, 163 have lower stored energies for the antiparallel initialized branch, while 123 have lower energies for the parallel initialized branch. These initialization labels do not independently establish the converged magnetic ordering; 157 branch energy differences are smaller than $10^{-5}$~eV per cell. We find no statistically significant relationship between unit cell volume and the lower energy initialization branch. The point biserial correlation between volume and initialization branch is 0.0788 ($p=0.184$), while a Mann-Whitney test gives $p=0.266$~\cite{Whitney_U_test}. Similarly, unit cell volume shows negligible correlations with either the magnitude of the [001] combined projected moment ($\rho=-0.009$, $p=0.884$; 285 compounds) or $K_1$ ($\rho=-0.090$, $p=0.130$) based on Spearman rank correlations (Fig.~\ref{fig:correlations}).

\begin{figure*}[t]\centering
\includegraphics[width=0.85\textwidth]{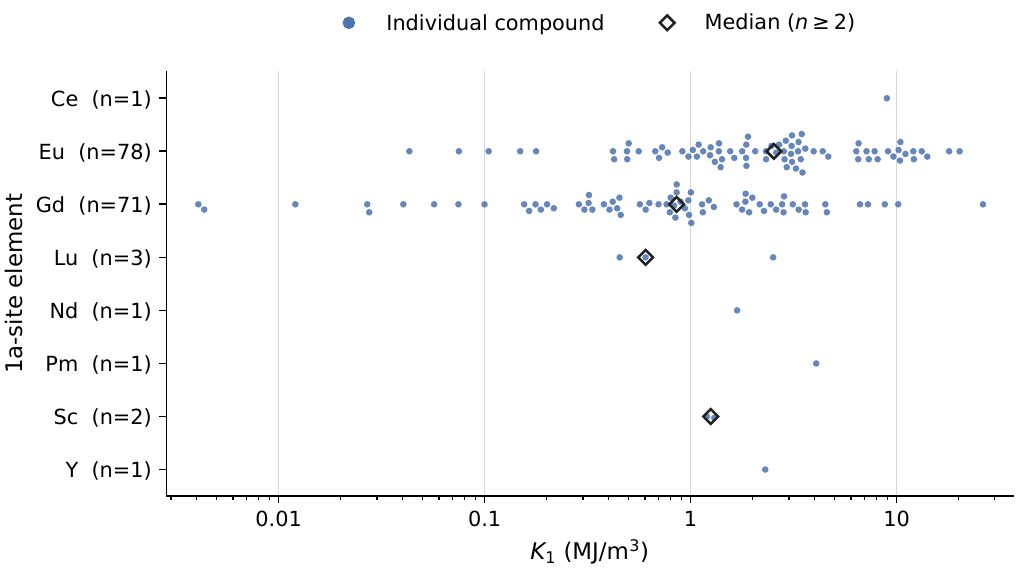}
\caption{Positive magnetic anisotropy energy densities $K_1$ for 158 CaCu$_5$-derived compounds, grouped by the element
occupying the $1a$ site. Sr is omitted. Each blue point represents one compound. The vertical offsets separate overlapping points and have no physical significance. Open diamonds indicate group medians for groups containing at least two compounds, and $n$ denotes the number of compounds in each group. The horizontal axis is logarithmic.}
\label{fig:re-mae}\end{figure*}

Among the 286 compounds in the final analysis set, 159 have positive $K_1$, 121 favor the tested in-plane directions ($K_1<0$), and six are effectively degenerate at the stored numerical precision. A Kruskal-Wallis test indicates differences in the positive-$K_1$ distributions among $1a$ site groups excluding Sr ($p=0.00052$)~\cite{Kruskal01121952,corder2009nonparametric}. However, the $1a$ site groups contain different numbers of compounds, so these differences should be interpreted with caution. The positive $K_1$ distributions for the different $1a$ site elements are shown in Figure~\ref{fig:re-mae}. Site-resolved analysis further indicates that no individual transition-metal element consistently produces strong anisotropy at both the $3g$ and $2c$ sites.

\begin{figure*}[t]\centering
\includegraphics[width=0.94\textwidth]{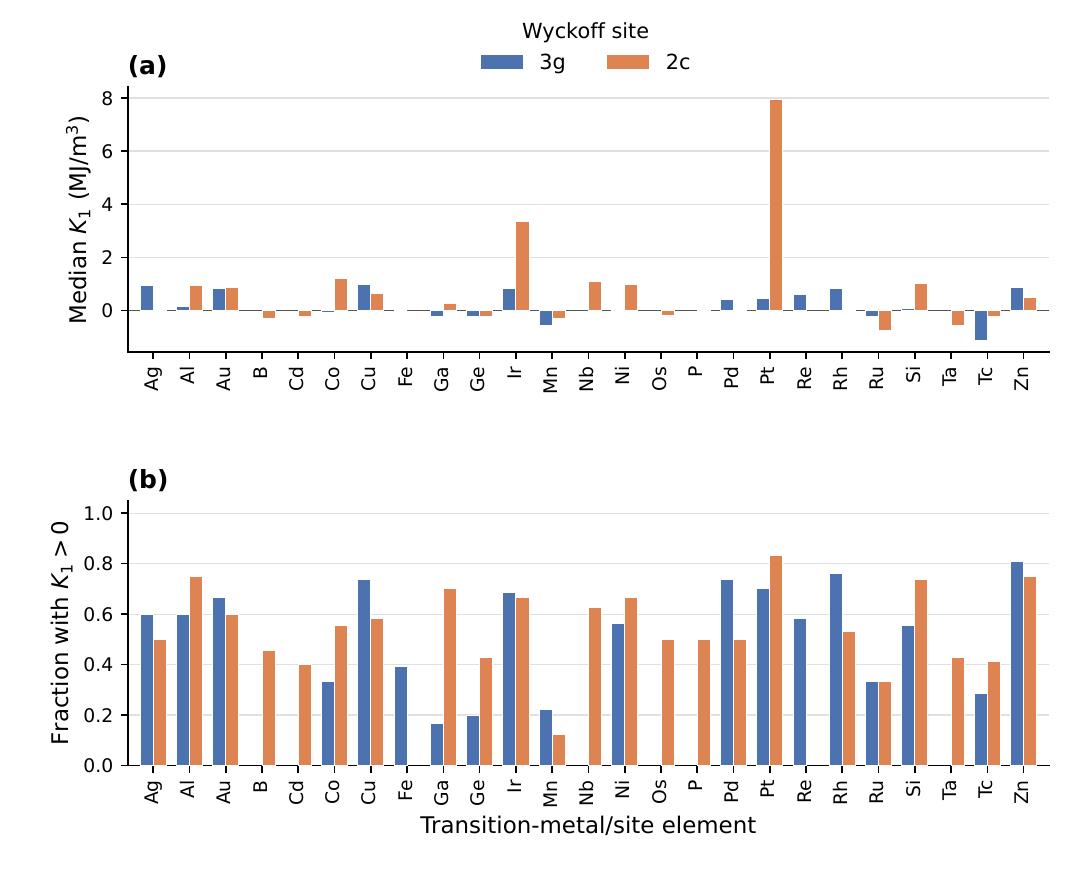}
\caption{For elements represented by at least five structures at a given site: (a) median magnetic anisotropy energy ($K_1$) and (b) fraction of compounds exhibiting positive $K_1$. Missing bars indicate fewer than five compounds at that site.}\label{fig:site-mae}\end{figure*}

To further examine the role of transition-metal chemistry, Figure~\ref{fig:site-mae} summarizes the median $K_1$ and fraction of compounds with positive $K_1$ for elements represented by at least five structures at a given site. The results illustrate site dependent associations between elemental occupancy and magnetic anisotropy within the retained dataset, rather than isolated causal contributions from individual elements.

The ten compounds with the largest positive $K_1$ values are listed in Table~\ref{tab:top}. GdTc$_3$Ir$_2$ has the largest calculated anisotropy, with $K_1=26.27$~MJ/m$^3$, followed by several Eu containing compounds with $K_1$ values between 10 and 20~MJ/m$^3$. The corresponding [001] magnetic moments range from 1.66 to 9.12~$\mu_B$/f.u., and both parallel and antiparallel initialized branches occur among the highest anisotropy compounds.

\begin{table}[t]
\caption{Ten systems with the largest positive axial anisotropy energy  $K_1$. Combined projected spin + orbital magnetic moments are reported as magnitudes along [001]. FM and FI denote the ferromagnetic and ferrimagnetic branches, not independently verified magnetic ordering.}
\label{tab:top}
\centering
\begin{tabular}{lccc}
\hline\hline
System & $K_1$ (MJ/m$^3$) & $|m_{\rm tot}^{\rm proj}|$ ($\mu_B$/f.u.) & Init. \\
\hline
GdTc$_3$Ir$_2$  & 26.27 & 1.66 & FI \\
EuPd$_3$Pt$_2$  & 20.23 & 6.12 & FM \\
EuPt$_3$Rh$_2$  & 17.99 & 5.54 & FI \\
EuZn$_3$Ir$_2$  & 14.09 & 5.19 & FM \\
EuTa$_2$Rh$_3$  & 13.25 & 6.17 & FM \\
EuRe$_3$Tc$_2$  & 12.13 & 7.41 & FM \\
EuSi$_2$Ir$_3$  & 12.05 & 6.33 & FI \\
EuSi$_2$Rh$_3$  & 11.05 & 6.48 & FM \\
EuTa$_3$Pt$_2$  & 10.44 & 9.12 & FM \\
EuPd$_2$Pt$_3$  & 10.39 & 6.01 & FI \\
\hline\hline
\end{tabular}
\end{table}

This ranking is based on anisotropy alone, not overall practical permanent magnet performance. Phase stability, elemental practicality, Curie temperature, and coercivity were not combined into a final performance ranking.

Although the screened dataset is dominated by RE containing compositions, the workflow also retained LiC$_2$Mn$_3$, MgP$_2$Mn$_3$, and SrNi$_2$Fe$_3$, all of which satisfied the anisotropy quality criteria. LiC$_2$Mn$_3$ and MgP$_2$Mn$_3$ exhibit negative $K_1$, whereas SrNi$_2$Fe$_3$ exhibits positive $K_1$ of approximately 1.82~MJ/m$^3$ and a total projected [001] magnetic moment of 10.67~$\mu_B$/f.u. SrNi$_2$Fe$_3$ remains in the final dataset despite being omitted from Fig.~\ref{fig:re-mae}. These results demonstrate that the workflow can identify magnetic candidates beyond RE substitution at the $1a$ site, although their synthesis feasibility and magnetic ground states require further investigation.

\subsection{Selected Ce-Fe based candidates}

Ce-Fe based magnetic materials have long attracted interest in the permanent magnet community~\cite{snv6-rd3c,m6nl-yxbv, PhysRevApplied.17.064020,RAJA2025173294}. Incorporating Fe can increase magnetization and reduce reliance on Co, although these benefits must be balanced against changes in phase stability and magnetic anisotropy. For example, Landa \textit{et al.}~\cite{Landa2020} investigated Fe- and Ni-containing CaCu$_5$ type compounds derived from YCo$_5$, demonstrating the potential of site selective substitution to combine favorable magnetic properties with negative formation energies. Experimental studies of substituted CeCo$_5$ further show that suitable microstructures are essential for developing coercivity~\cite{Lamichhane2019}. Here, we examine two Ce-Fe based compositions selected by our screening workflow, CeNi$_2$Fe$_3$, which exhibits large positive axial magnetic anisotropy in the selected calculations, and CeMn$_2$Fe$_3$, which combines sizable magnetization with planar anisotropy.

These detailed WIEN2k results are distinct from the VASP screening dataset. The following case studies illustrate two complementary limitations of candidate selection: favorable magnetic properties do not establish thermodynamic accessibility, and substantial magnetization does not necessarily imply easy-axis anisotropy.

\subsubsection{CeNi$_2$Fe$_3$}
\label{sub:CeNi2Fe3}

Our ML screening retained CeNi$_2$Fe$_3$ as a candidate for detailed first-principles investigation. We evaluated its magnetic properties using the all-electron WIEN2k code within DFT + $U$, including SOC. The computed magnetic moment is $\approx 0.9$~T and $K_1$ is approximately 15.4~MJ/m$^3$, favoring [001] over [100]. The details regarding the calculations are provided in Supplemental Material Section~III.1.1 and Table~S2, with the elemental reference energies in Section~III.1.2 and Table~S3 and the formation-energy and decomposition expressions in Eqs.~S1--S2~\cite{supp}. 

The Fe-Ni combination has precedent in related CaCu$_5$-type compounds. Landa \textit{et al.}~\cite{Landa2020} investigated YFe$_3$(Ni$_{1-x}$Co$_x$)$_2$, with Fe occupying the $3g$ sites and Ni/Co occupying the $2c$ sites, and reported negative formation energies over a broad composition range. Their calculations also predicted positive $K_1$ for YFe$_3$Ni$_2$. These results provide motivation for exploring analogous Fe--Ni compositions containing Ce, while recognizing that the stability and magnetic properties of a given composition may differ substantially with the choice of rare-earth element.

\begin{figure*}[tbp]
\centering
\includegraphics[width=0.85\textwidth]
{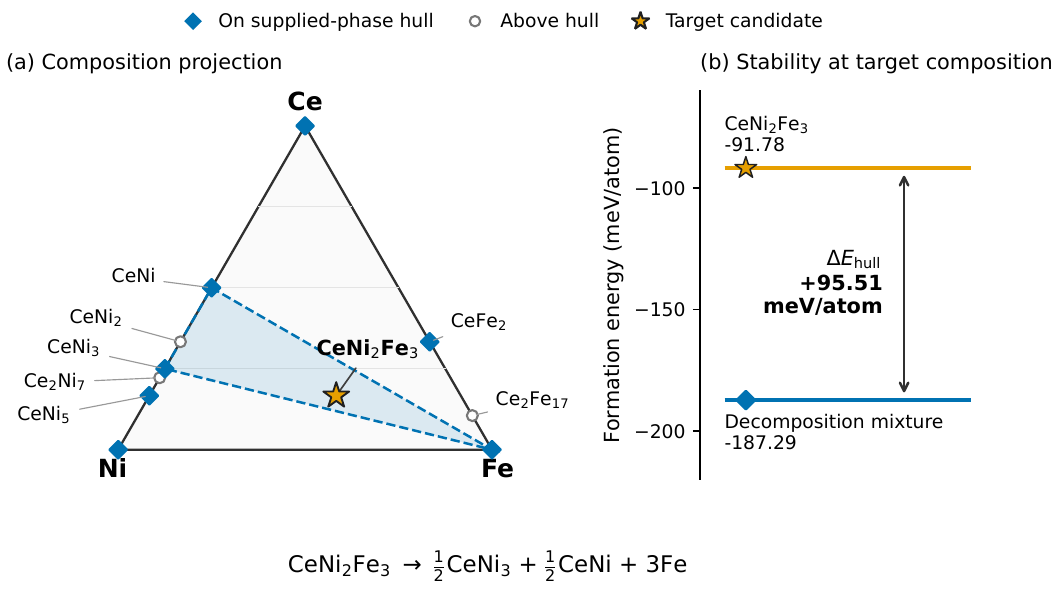}
\caption{Ce-Ni-Fe stability analysis using the supplied WIEN2k phase energies. (a) Composition projection showing phases on the supplied-phase convex hull (blue diamonds), phases above the hull (open circles), and ordered CeNi$_2$Fe$_3$ (gold star). The shaded region and dashed boundary highlight the CeNi$_3$-CeNi-Fe decomposition triangle; point positions represent atomic compositions, not energies. (b) Formation energies at the same overall CeNi$_2$Fe$_3$ composition. The candidate lies 95.51~meV/atom above the decomposition mixture $\frac{1}{2}$CeNi$_3$ + $\frac{1}{2}$CeNi + 3Fe. The hull is restricted to the supplied phase set and remains subject to reference-phase convergence limitations.}
\label{fig:E_hull_CeNi2Fe3}
\end{figure*}

Experimental work by Pourarian \textit{et al.}~\cite{POURARIAN1986111} provides a more direct precedent. Fe substitution transforms Pauli-paramagnetic CeNi$_5$ into a ferromagnetic material, with Fe occupying both the $2c$ and $3g$ sites. Single-phase CeNi$_{5-x}$Fe$_x$ alloys were reported for $x=0.5$, 1.0, 1.5, and 1.75. However, these compositions contain substantially less Fe than CeNi$_2$Fe$_3$ ($x=3$), so their synthesis does not establish the accessibility of the ordered composition considered here.

To examine its thermodynamic stability, we calculated the elemental formation energy per atom as

\begin{equation}
\label{eq:CeNi2Fe3_formation}
E_f =
\frac{
E_{\mathrm{tot}}(\mathrm{CeNi_2Fe_3})
-E_{\mathrm{Ce}}
-2E_{\mathrm{Ni}}
-3E_{\mathrm{Fe}}
}{6},
\end{equation}

\noindent
where $E_{\mathrm{tot}}(\mathrm{CeNi_2Fe_3})$ is the energy per formula unit and $E_{\mathrm{Ce}}$, $E_{\mathrm{Ni}}$, and $E_{\mathrm{Fe}}$ are the corresponding elemental reference energies per atom. The resulting value of $-91.8$~meV/atom indicates an energetic preference relative to the constituent elements. Nevertheless, the compound lies approximately 95.5~meV/atom above the convex hull constructed from the supplied Ce-Ni-Fe phase energies. Figure~\ref{fig:E_hull_CeNi2Fe3}(a) highlights the CeNi$_3$-CeNi-Fe decomposition triangle containing the candidate composition. The corresponding decomposition per formula unit is

\[
\mathrm{CeNi_2Fe_3}
\rightarrow
\frac{1}{2}\mathrm{CeNi_3}
+\frac{1}{2}\mathrm{CeNi}
+3\mathrm{Fe}.
\]

As shown in Figure~\ref{fig:E_hull_CeNi2Fe3}(b), this mixture has a formation energy of approximately $-187.3$~meV/atom at the same overall composition, compared with $-91.8$~meV/atom for CeNi$_2$Fe$_3$. The ordered compound is therefore predicted to be metastable relative to these competing phases at zero temperature. This assessment is restricted to the supplied phase set and remains subject to convergence limitations in the reference-phase calculations.

Chemical disorder and finite temperature contributions can modify this stability assessment, but their effects must be evaluated for both the candidate and its competing phases. The experimentally reported CeNi$_{5-x}$Fe$_x$ alloys therefore provide motivation for further study, rather than proof of entropy stabilization at $x=3$. CeNi$_2$Fe$_3$ remains an interesting metastable magnetic candidate because of its sizable magnetic moment and large positive axial anisotropy in the selected calculations. Its experimental accessibility, magnetic hardness, and anisotropy convergence require further assessment. Moreover, Landa \textit{et al.}~\cite{Landa2020} found substantial method dependence in the calculated anisotropy of related CaCu$_5$-type compounds, reinforcing the need to interpret the predicted anisotropy quantitatively with caution.

\subsubsection{CeMn$_2$Fe$_3$}

CeMn$_2$Fe$_3$ is noteworthy for retaining a substantial predicted saturation magnetic polarization, $\mu_0M_s \approx 1.3$~T, despite Mn occupying 40\% of the transition metal sites. In contrast, high-field measurements on Ce-Fe-Mn based magnets such as Ce$_2$Fe$_{17-x}$Mn$_x$~\cite{Kuchin2007CeFeMn} indicate that Mn moments couple antiparallel to Fe, reducing the net magnetization. In our WIEN2k calculations, the Mn and Fe moments are parallel, whereas the Ce moment is antiparallel to both. This result highlights a potentially favorable magnetic configuration in the ordered CaCu$_5$-type structure, rather than a general absence of high-magnetization Fe-Mn compounds.

Ce-Fe-Mn intermetallics have previously been synthesized in other structure families. Shtender \textit{et al.}~\cite{Shtender2021} identified Ce$_3$(Fe$_{0.638}$Mn$_{0.362}$)$_{29}$ in the monoclinic Nd$_3$(Fe,Ti)$_{29}$ structure and reported magnetization consistent with ferrimagnetic ordering above 400~K. They also identified Ce(Fe,Mn)$_{12}$ and Ce$_2$(Fe,Mn)$_{17}$ phases. These observations establish experimental precedents for Ce-Fe-Mn chemistry, but do not demonstrate permanent magnet performance or the synthesis of the ordered CaCu$_5$ type CeMn$_2$Fe$_3$ structure studied here. The suggested ferrimagnetic ordering of the 3:29 phase also does not independently establish the relative alignment of its Fe and Mn moments.

Our WIEN2k GGA + $U$ calculations including SOC give $K_1 =-0.82$~MJ/m$^3$ for CeMn$_2$Fe$_3$, corresponding to planar anisotropy. Consequently, despite its sizable magnetization, this composition does not satisfy the easy axis criterion used to identify uniaxial permanent magnet candidates in the present workflow. It instead provides a useful example of why magnetization alone is insufficient for candidate selection.

The elemental formation energy, evaluated using the normalization in Eq.~\ref{eq:CeNi2Fe3_formation} with Mn replacing Ni, is $-40$~meV/atom. This indicates an energetic preference relative to the constituent elements, but does not establish stability against competing Ce-Mn-Fe compounds. A consistent convex-hull assessment was not available for this composition in the present study. The experimentally reported Ce-Fe-Mn phases provide relevant competitors for such an assessment, and a hull distance remains necessary to evaluate the thermodynamic viability of the proposed CaCu$_5$-type structure.

\section{Discussion\label{sec:discussion}}

In this work, we developed a machine learning assisted first-principles workflow for screening CaCu$_5$-type magnetic materials. Starting from 84,564 unique candidate compositions generated by site-selective substitution, the combined machine learning and materials data screening reduced the chemical space to 439 candidates for density functional theory calculations. Usable spin-orbit coupled calculations were obtained for 407 compounds (see Supplemental Material Section~III.2.3 and Table~S5), and subsequent quality-control criteria yielded 286 structures for analysis, including 159 with positive magnetocrystalline anisotropy energy favoring the [001] direction.

The workflow substantially reduces the number of compositions requiring computationally expensive first-principles calculations, retaining only 0.708\% of the jointly evaluable candidate pool. While this reduction provides a direct measure of the computational benefit, it should not be interpreted as a measured wall-time or CPU-hour speedup. Analysis of the resulting dataset reveals composition- and site-dependent variations in magnetic anisotropy that are not explained by unit-cell volume alone. Several candidates exhibit sizable positive anisotropy and large calculated magnetic moments, providing targets for further computational and experimental investigation. More broadly, this work demonstrates how existing machine learning models can be combined with first-principles calculations to systematically reduce large chemical spaces and accelerate the search for candidate permanent-magnet materials.

\section*{Acknowledgements}

This work is supported by the Critical Materials Innovation Hub, an Energy Innovation Hub funded by the US Department of Energy, Office of Critical Minerals and Energy Innovation (CMEI), Advanced Materials and Manufacturing Technologies Office (AMMTO). This research used resources of the Compute and Data Environment for Science (CADES) at the Oak Ridge National Laboratory, which is supported by the Office of Science of the U.S. Department of Energy under Contract No. DEAC0500OR22725. We would also like to acknowledge the Critical Materials Innovation Hub, Ames National Laboratory, U.S. Department of Energy, Ames, IA, USA. This manuscript has been authored by UT-Battelle, LLC under Contract No. DEAC0500OR22725 with the U.S. Department of Energy. The United States Government retains and the publisher, by accepting the article for publication, acknowledges that the United States Government retains a nonexclusive, paid-up, irrevocable, world-wide license to publish or reproduce the published form of this manuscript, or allow others to do so, for United States Government purposes. The Department of Energy will provide public access to these results of federally sponsored research in accordance with the DOE Public Access Plan.

\section*{Author Contributions}
N.P. and S.W. have made equal contributions to this work.

\section*{Data Availability Statement}
Raw data were generated at the Oak Ridge National Laboratory large scale facility. Derived data supporting the findings of this study are available from the corresponding author upon reasonable request.

\bibliography{references} 
\end{document}


\begin{titlepage}
\thispagestyle{empty}
\centering
\vspace*{1.2in}
{\Large\bfseries Data driven screening of CaCu$_5$-type magnetic structures\par}
\vspace{0.6in}
{\large Nabaraj Pokhrel$^{1}$, Sheila Whitman$^{2}$, and David S. Parker$^{1}$\par}
\vspace{0.3in}
{$^{1}$Materials Science and Technology Division, Oak Ridge National Laboratory,\\
Oak Ridge, TN 37831, USA\par}
\vspace{0.2in}
{$^{2}$Department of Applied Mathematics, University of Arizona,\\
Arizona 85721, USA\par}
\vspace{0.75in}
{\Large Supplemental Material\par}
\vfill
\end{titlepage}
\setcounter{page}{2}

\section{CaCu$_5$ structure and chemical space}\label{sec:generation}
Candidates were generated by substitution in a six-atom CaCu$_5$-type cell with nominal composition $AB_3C_2$. The $A$, $B$, and $C$ substitutions occupy $1a$, $3g$, and $2c$, respectively. Identical $B$ and $C$ elements give $AX_5$. The generator's atom sequence is $A,C,C,B,B,B$. Archive formula element order is preserved and is not by itself a site-occupancy specification.

The initial lattice vectors, in \AA, were
\[
\begin{pmatrix}
5.0456280708 & 0 & 0\\
-2.5228140354 & 4.3696420874 & 0\\
0 & 0 & 4.0173730850
\end{pmatrix}.
\]
Thus, $a=b=5.0456280708$~\AA, $c=4.0173730850$~\AA, with initial angles $90^\circ$, $90^\circ$, $120^\circ$. All compositions began from this same cell; relaxation could change lattice and atomic positions.
Starting fractional coordinates, rounded to ideal Wyckoff positions, are given below. The generator uses numerically approximated thirds; original atom order is retained.
\begin{center}
\begin{tabular}{cccrrr}\toprule
Atom & Substitution & Wyckoff site & $x$ & $y$ & $z$\\\midrule
1 & $A$ & $1a$ & 0 & 0 & 0\\
2 & $C$ & $2c$ & $1/3$ & $2/3$ & 0\\
3 & $C$ & $2c$ & $2/3$ & $1/3$ & 0\\
4 & $B$ & $3g$ & 0 & $1/2$ & $1/2$\\
5 & $B$ & $3g$ & $1/2$ & $1/2$ & $1/2$\\
6 & $B$ & $3g$ & $1/2$ & 0 & $1/2$\\\bottomrule
\end{tabular}
\end{center}

\subsection{Element-selection and model-support restrictions}\label{sec:support}
The generator defines these pools:
\begin{itemize}
\item RE/Sc/Y: Lu, La, Ce, Pr, Nd, Pm, Sm, Eu, Gd, Tb, Dy, Ho, Er, Tm, Yb, Sc, Y.
\item Alkali and alkaline-earth elements: Li, Na, K, Rb, Cs, Fr; Be, Mg, Ca, Sr, Ba, Ra.
\item Transition metals: Ti, V, Cr, Mn, Fe, Co, Ni, Cu, Zn; Zr, Nb, Mo, Tc, Ru, Rh, Pd, Ag, Cd; Hf, Ta, W, Re, Os, Ir, Pt, Au, Hg.
\item Other elements and metalloids: Al, Ga, In, Sn, Tl, Pb, Bi; B, Si, Ge, As, Sb, Te.
\item Additional nonmetal and noble-gas substitutions: C, N, O, F, Ne, P, S, Cl, Ar, Se, Br, Kr, I, Xe.
\end{itemize}

These are generator pools, not a claim that every element is supported by all models. Candidates must be evaluable by both pretrained graph networks and the descriptor/regressor pipeline. We excluded the unsupported compositions. 

\subsection{Screening models, thresholds, and accounting}\label{sec:screening}
M3GNet-MP-2021.2.8-DIRECT-PES was used for relaxation~\cite{M3GNET,qi2024robust}. MEGNet-MP-2019.4.1-BandGap-mfi predicted band gaps with model state attribute 0~\cite{Chen2019,chen2021learning}. A random forest was used to predict the magnetization. 

For stability screening, the relaxation energy per atom was multiplied by atom count to construct a candidate entry, which was inserted into a phase diagram containing Materials Project GGA/GGA+$U$ entries for that chemical system~\cite{materials_project}. Formation energy and hull distance were then calculated. Candidate energies originate from the ML potential, whereas competitor energies originate from database calculations. These mixed-source quantities are screening estimates, not rigorously compatible DFT formation energies or hull distances.
The four criteria were evaluated independently on the common pool of 62,017 jointly evaluable candidates where 52,864 satisfied the metallicity threshold (predicted band gap $\leq0$), 5,984 had $\Delta E_{\rm hull}\leq50$~meV/atom, 34,119 had $E_f<0$, and 3,102 had predicted magnetization $\geq0.75$~T. Their joint intersection contained 439 candidates. These are independent passing counts, not sequential attrition counts. The initial generator produced 84,564 unique identifiers.

The $1a$ counts among 439 are Gd 176, Eu 173, Pm 12, Tm 9, Er 8, Tb 8, Y 8, Sm 7, Dy 6, Lu 6, Nd 6, Ce 5, Pr 4, Ho 3, La 3, Sc 2, Li 1, Mg 1, Sr 1.

\section{Magnetization model evaluation}\label{sec:features}

Features include $a$, $b$, $c$, atom count, elemental fractions, Magpie composition statistics, and volume in bohr$^3$, using Matminer composition descriptors~\cite{ward2018matminer}. The elemental-fraction columns are explicitly:

\begin{quote}\small
H, He, Li, Be, B, C, N, O, F, Ne, Na, Mg, Al, Si, P, S, Cl, Ar, K, Ca, Sc, Ti, V, Cr, Mn, Fe, Co, Ni, Cu, Zn, Ga, Ge, As, Se, Br, Kr, Rb, Sr, Y, Zr, Nb, Mo, Tc, Ru, Rh, Pd, Ag, Cd, In, Sn, Sb, Te, I, Xe, Cs, Ba, La, Ce, Pr, Nd, Pm, Sm, Eu, Gd, Tb, Dy, Ho, Er, Tm, Yb, Lu, Hf, Ta, W, Re, Os, Ir, Pt, Au, Hg, Tl, Pb, Bi, Ac, Th, Pa, U, Np, Pu.
\end{quote}

This is a specified list, not every element from H through Pu. Minimum, maximum, range, mean, average deviation, and mode are supplied for each of these 22 Magpie properties (132 columns):
\begin{quote}\small
Number, MendeleevNumber, AtomicWeight, MeltingT, Column, Row, CovalentRadius, Electronegativity, NsValence, NpValence, NdValence, NfValence, NValence, NsUnfilled, NpUnfilled, NdUnfilled, NfUnfilled, NUnfilled, GSvolume\_pa, GSbandgap, GSmagmom, SpaceGroupNumber.
\end{quote}

Randomized search cross validation tested 100, 200, and 300 trees with maximum depths of None, 10, 20, and 30, minimum split sizes of 2, 5, and 10, minimum leaf sizes of 1, 2, and 4, and maximum-feature settings of auto, square root, and log$_2$. The selected model used 300 trees, a maximum depth of 30, a minimum split size of 10, a minimum leaf size of 4, and square-root feature sampling.

\begin{table}[H]\centering
\caption{Reported magnetization metrics. Initial comparisons are fivefold means. The optimized test and prototype diagnostic are different evaluations.}\label{tab:metrics}
\begin{tabular}{lrrr}\toprule
Evaluation & MAE (T) & RMSE (T) & $R^2$\\\midrule
Initial linear regression & 0.1249 & 0.1986 & 0.6875\\
Initial ridge regression & 0.1249 & 0.1986 & 0.6875\\
Initial SVR (RBF kernel) & 0.0897 & 0.1575 & 0.8036\\
Initial random forest & 0.0597 & 0.1405 & 0.8437\\
Optimized random-forest test & 0.0664 & 0.1414 & 0.8361\\
CaCu$_5$ diagnostic & 0.0399 & 0.1148 & 0.9409\\\bottomrule
\end{tabular}\end{table}

The 80:20 train-test split has 123,893 training and 30,974 test structures. The final test MSE is 0.0200~T$^2$. Initial fold standard deviations are unavailable. Selection was not nested, and the known-prototype set was not excluded from general train/test data. Metrics are internal evaluations, not an external benchmark. Of 204 known CaCu$_5$ structures, 25 were unsupported, leaving 179. MEGNet classified 97.8\% of known metallic cases as metallic. This should be taken as a positive control as no known nonmetallic controls were available.

\section{Chemical-space analysis and Ce-based case studies}\label{sec:analysis}
\subsection{WIEN2k calculations for CeNi$_2$Fe$_3$}\label{sec:wien}
The all-electron case study is separate from the VASP screening catalog. 

\subsubsection{Structure and numerical settings}
The six atom cell contains one CeNi$_2$Fe$_3$ formula unit, with Ce at $1a$, Fe at $3g$, and Ni at $2c$. The lattice parameters are $a=b=5.241$~\AA\ and $c=4.218$~\AA, with angles $90^\circ$, $90^\circ$, and $120^\circ$. The SCF cell volume is $100.33755872$~\AA$^3$.
\begin{table}[H]\centering

\caption{Saved WIEN2k settings for the selected CeNi$_2$Fe$_3$ baseline calculations.}\label{tab:wien-settings}
\begin{tabular}{p{2.15in}p{3.65in}}\toprule
Quantity & Setting\\\midrule
Code and functional & WIEN2k 24.1; PBE\\
Hubbard correction &  $U = 3$ eV, $J=0$ applied to Ce $4f$\\
Muffin-tin radii & Ce: 2.50 bohr; Fe and Ni: 2.34 bohr\\
Basis and angular cutoff & $R_{\rm MT}K_{\max}=7.0$; wavefunction angular cutoff 10\\
Charge-density cutoff & $G_{\max}=12$ bohr$^{-1}$\\
Integration & Tetrahedron method\\
Selected SOC mesh & $9\times9\times10$ divisions; 150 final SCF $k$ points\\
Charge criterion & Requested atomic charge convergence: $10^{-4}$
\\\bottomrule
\end{tabular}\end{table}

\subsubsection{Formation energy and supplied-phase convex hull}
The formation energy calculation uses the scalar relativistic CeNi$_2$Fe$_3$ energy $-31450.60752356$~Ry/f.u. and the elemental references in Table~\ref{tab:wien-refs}. Although one archive folder is named \texttt{VASP\_convex\_hull}, the supplied energies used here are from WIEN2k, not a mixture of absolute WIEN2k and VASP energies.
 
 \begin{table}[H]\centering
\caption{Elemental reference energies used in the WIEN2k formation-energy calculation, in Ry/atom.}\label{tab:wien-refs}
\begin{tabular}{lr}\toprule
Reference & Energy\\\midrule
fcc Ce & $-17730.64842769$\\
bcc Fe & $-2545.56026341$\\
fcc Ni & $-3041.61891549$\\\bottomrule
\end{tabular}\end{table}
With normalization per atom,
\begin{equation}
\label{eq:formation}
E_f=\frac{E_{\rm tot}(\mathrm{CeNi_2Fe_3})-E_{\rm Ce}-2E_{\rm Ni}-3E_{\rm Fe}}{6}
=-91.780967~\mathrm{meV/atom}.
\end{equation}

The Ce reference uses the same Ce $4f$ correction and double counting convention. The phase set consists of Ce, Fe, Ni, CeNi, CeNi$_2$, CeNi$_3$, CeNi$_5$, Ce$_2$Ni$_7$, CeFe$_2$, and Ce$_2$Fe$_{17}$. Minimization over this finite set gives a hull distance of $+95.510002$~meV/atom, with decomposition

\begin{equation}
\mathrm{CeNi_2Fe_3}\rightarrow
\tfrac12\mathrm{CeNi_3}+\tfrac12\mathrm{CeNi}+3\mathrm{Fe}.
\end{equation}

\subsubsection{Relation to the VASP screening calculations}
The VASP and WIEN2k cases must not be interchanged. The results from VASP calculations encouraged us to further test these results using WIEN2k.

\subsection{VASP settings, initializations, and quality control}\label{sec:vasp}
Calculations were performed using VASP~\cite{kresse95} with the PBE functional~\cite{PBE_functional}, PAW potentials~\cite{PAW_method}, a plane-wave cutoff of 500~eV, and a $\Gamma$-centered $4\times4\times4$ $k$-point mesh. The electronic and force convergence thresholds used were $10^{-7}$~eV and 0.01~eV/\AA, respectively. Parallel and antiparallel initial alignments between the RE and transition-metal moments were labelled \texttt{FM} and \texttt{FI}, respectively. The input generator assigned initial moments of $8\,\mu_B$ to RE atoms, $2\,\mu_B$ to transition metals, and $1\,\mu_B$ to other listed species, with the RE alignment reversed for the antiparallel initialization. 

\subsubsection{Hubbard parameters}
Hubbard corrections were applied using the Dudarev approach with \texttt{LDAUTYPE=2} and \texttt{LDAUJ=0}. Table~\ref{tab:u} lists the nonzero $U_{\rm eff}=U-J$ values used by the input generator for the RE $4f$ states and the listed transition-metal $d$ states, including Sc and Y. All other species were assigned zero $U$. The values in the completed [001] calculation inputs were checked against the generator, although some elements listed in the table do not appear in those calculations. The parameters follow Topsakal and Wentzcovitch~\cite{topsakal2014accurate} and the Materials Project conventions~\cite{materials_project}.

\begin{longtable}{lrl}
\caption{Nonzero Hubbard parameters in the archived generator.}\label{tab:u}\\
\toprule Element & $U_{\rm eff}$ (eV) & Channel\\\midrule\endfirsthead
\toprule Element & $U_{\rm eff}$ (eV) & Channel\\\midrule\endhead
\bottomrule\endfoot
Ag & 2.25 & $d$ \\
Au & 1.12 & $d$ \\
Cd & 0.24 & $d$ \\
Ce & 2.5 & 4$f$ \\
Co & 3.32 & $d$ \\
Cr & 3.7 & $d$ \\
Cu & 7.77 & $d$ \\
Dy & 5 & 4$f$ \\
Er & 4.2 & 4$f$ \\
Eu & 3 & 4$f$ \\
Fe & 5.3 & $d$ \\
Gd & 4.6 & 4$f$ \\
Hf & 3.52 & $d$ \\
Hg & 0.52 & $d$ \\
Ho & 4.9 & 4$f$ \\
Ir & 1.9 & $d$ \\
La & 5.5 & 4$f$ \\
Lu & 5.5 & 4$f$ \\
Mn & 3.9 & $d$ \\
Mo & 4.38 & $d$ \\
Nb & 0.54 & $d$ \\
Nd & 3.1 & 4$f$ \\
Ni & 6.2 & $d$ \\
Os & 1.91 & $d$ \\
Pd & 3.76 & $d$ \\
Pm & 3.4 & 4$f$ \\
Pr & 4 & 4$f$ \\
Pt & 1.55 & $d$ \\
Re & 0.62 & $d$ \\
Rh & 1.53 & $d$ \\
Ru & 3 & $d$ \\
Sc & 1.92 & $d$ \\
Sm & 3.3 & 4$f$ \\
Ta & 3.73 & $d$ \\
Tb & 5 & 4$f$ \\
Tc & 2.95 & $d$ \\
Ti & 4.91 & $d$ \\
Tm & 4.8 & 4$f$ \\
V & 3.25 & $d$ \\
W & 6.2 & $d$ \\
Y & 4.3 & $d$ \\
Yb & 3 & 4$f$ \\
Zn & 1.92 & $d$ \\
Zr & 4.2 & $d$ \\
\end{longtable}

\subsubsection{SOC energies and quality filters}
SOC used fixed charge density along [001], [100], [010] crystalline directions. All 408 inputs specify 500 eV cutoff, $10^{-7}$ eV electronic threshold, and 0.02 eV smearing.

The screening magnetic anisotropy energy is

\begin{equation}
K_1=\frac{E_{[100]}-E_{[001]}}{V},\qquad
1\ {\rm eV}/\text{\AA}^3=160217.6634\ {\rm MJ/m^3}.
\end{equation}

The retained set satisfies

\begin{equation}
|E_{[100]}-E_{[010]}|/V\leq0.10\ {\rm MJ/m^3},
\qquad |K_1|\leq30\ {\rm MJ/m^3}.
\end{equation}

There are 159 positive, 121 negative, six zero values at stored precision. These filters test selected numerical conditions, not full convergence, phonon stability, thermodynamics, or magnetic ground state validity. In-plane differences could reflect uncertainty or symmetry lowering.

\subsubsection{Failures and accounting gaps}
The following set of structures failed to converge for the given set of input parameters.

\begin{longtable}{ll}
\caption{Documented failures. Here, `both' refers to FM and FI together.}\label{tab:failures}\\
\toprule Composition & Recorded status\\\midrule\endfirsthead
\toprule Composition & Recorded status\\\midrule\endhead
\bottomrule\endfoot
ErFe$_2$Co$_3$ & FI failed\\
EuW$_2$Ru$_3$ & Both failed\\
EuW$_2$Tc$_3$ & Both failed\\
GdAs$_2$Rh$_3$ & FI failed\\
GdCo$_2$Fe$_3$ & FI failed\\
GdCo$_2$Mn$_3$ & FI failed\\
GdIr$_2$W$_3$ & Both failed\\
GdTc$_2$W$_3$ & Both failed\\
GdW$_2$Tc$_3$ & Both failed\\
PrNi$_2$Fe$_3$ & FI failed\\
SmCo$_2$Fe$_3$ & Both failed\\
SmCo$_5$ & Both failed\\
SmFe$_2$Co$_3$ & Both failed\\
SmMn$_2$Co$_3$ & Both failed\\
SmNi$_2$Fe$_3$ & Both failed\\
SmRu$_2$Fe$_3$ & Both failed\\
SmTc$_2$Fe$_3$ & Both failed\\
TbCo$_2$Fe$_3$ & FI failed\\
TbFe$_5$ & FI failed\\
TbNi$_2$Fe$_3$ & FM failed\\
EuNe$_5$ & FI failed\\
EuP$_2$Mo$_3$ & FI failed\\
\end{longtable}

\subsection{Ce/Sm candidates retained by the ML screen}\label{sec:cesm}
Table~\ref{tab:cesm} lists five Ce and seven Sm candidates passing the joint ML screen, not necessarily DFT quality control.
\begin{table}[H]\centering
\caption{Ce/Sm compositions retained by the four-criterion ML screen.}\label{tab:cesm}
\begin{tabular}{p{0.6in}p{5.0in}}\toprule
$1a$ & Compositions\\\midrule
Ce & CeCo$_2$Fe$_3$, CeCo$_5$, CeFe$_5$, CeMn$_2$Fe$_3$, CeNi$_2$Fe$_3$.\\
Sm & SmCo$_2$Fe$_3$, SmCo$_5$, SmFe$_2$Co$_3$, SmMn$_2$Co$_3$, SmNi$_2$Fe$_3$, SmRu$_2$Fe$_3$, SmTc$_2$Fe$_3$.\\\bottomrule
\end{tabular}\end{table}

\subsection{Complete quality-filtered VASP catalog}\label{sec:catalog}
Table~\ref{tab:catalog} contains all 286 main-text quality-filtered VASP entries. Lattice lengths are norms of final direct-lattice vectors in [001] SOC output; $a=b$ is not imposed. Angles, coordinates, and site occupations are also needed to specify relaxed structures. All moments are projected [001] quantities in $\mu_B$/f.u.; signed $K_1$ is in MJ/m$^3$. Zeros mean equality at archived precision, not exact physical degeneracy. The CSV retains additional precision and source identifiers.
\begingroup\small\setlength{\tabcolsep}{4pt}
\renewcommand{\arraystretch}{1.10}
\begin{longtable}{rlrrrrrrr}
\caption{All 286 quality-filtered candidates. Lattice lengths: \AA; $K_1$: MJ/m$^3$; moments: $\mu_B$/f.u. A dash denotes an unavailable value.}\label{tab:catalog}\\
\toprule
No. & Composition & $a$ & $b$ & $c$ & $K_1$ & $m_s$ & $m_l$ & $m_{\rm tot}^{\rm proj}$\\\midrule\endfirsthead
\multicolumn{9}{l}{\textit{Table \thetable\ continued}}\\\toprule
No. & Composition & $a$ & $b$ & $c$ & $K_1$ & $m_s$ & $m_l$ & $m_{\rm tot}^{\rm proj}$\\\midrule\endhead
\midrule\multicolumn{9}{r}{\textit{Continued on next page}}\\\endfoot
\bottomrule\endlastfoot
1 & $\mathrm{CeCo_{5}}$ & 4.97822 & 4.97822 & 4.05284 & $+8.9709$ & 7.737 & 0.867 & 8.604 \\
2 & $\mathrm{CeMn_{2}Fe_{3}}$ & 5.50598 & 5.50735 & 4.70214 & $-0.8157$ & 16.883 & 0.231 & 17.114 \\
3 & $\mathrm{ErTc_{2}Fe_{3}}$ & 5.13468 & 5.14412 & 4.52261 & $-4.5917$ & 5.491 & \textemdash & \textemdash \\
4 & $\mathrm{EuAg_{2}Al_{3}}$ & 5.63051 & 5.63051 & 4.41447 & $-1.4729$ & 6.914 & -0.055 & 6.859 \\
5 & $\mathrm{EuAg_{2}Pd_{3}}$ & 5.54896 & 5.54896 & 4.40014 & $+0.0750$ & 6.643 & -0.093 & 6.550 \\
6 & $\mathrm{EuAg_{2}Zn_{3}}$ & 5.58833 & 5.58833 & 4.46305 & $+0.5599$ & 6.884 & -0.043 & 6.841 \\
7 & $\mathrm{EuAl_{2}Ag_{3}}$ & 5.78247 & 5.78247 & 4.30128 & $+1.8594$ & 6.976 & -0.051 & 6.925 \\
8 & $\mathrm{EuAl_{2}Cu_{3}}$ & 5.43515 & 5.43515 & 4.04808 & $+3.3428$ & 6.988 & -0.076 & 6.912 \\
9 & $\mathrm{EuAl_{2}Ni_{3}}$ & 5.34894 & 5.34895 & 3.72869 & $+3.0935$ & 6.113 & -0.247 & 5.866 \\
10 & $\mathrm{EuAl_{2}Os_{3}}$ & 5.43913 & 5.43913 & 4.14224 & $-26.6947$ & 4.232 & -0.417 & 3.815 \\
11 & $\mathrm{EuAl_{2}Pd_{3}}$ & 5.63969 & 5.63969 & 3.95093 & $+2.3227$ & 7.062 & -0.043 & 7.019 \\
12 & $\mathrm{EuAl_{2}Zn_{3}}$ & 5.58708 & 5.58708 & 4.21012 & $+3.0717$ & 6.955 & -0.098 & 6.857 \\
13 & $\mathrm{EuAl_{5}}$ & 5.64321 & 5.64321 & 4.44955 & $+1.0936$ & 6.825 & -0.062 & 6.763 \\
14 & $\mathrm{EuAs_{2}Cu_{3}}$ & 5.71638 & 5.71638 & 3.83019 & $+2.6871$ & 7.064 & -0.088 & 6.976 \\
15 & $\mathrm{EuAs_{2}Ni_{3}}$ & 5.67790 & 5.67792 & 3.76440 & $+4.6614$ & 4.560 & -0.173 & 4.387 \\
16 & $\mathrm{EuAu_{2}Al_{3}}$ & 5.64813 & 5.64813 & 4.31101 & $-0.6051$ & 6.825 & -0.086 & 6.739 \\
17 & $\mathrm{EuAu_{2}Pd_{3}}$ & 5.52474 & 5.52474 & 4.43954 & $+3.9578$ & 6.825 & -0.137 & 6.688 \\
18 & $\mathrm{EuB_{2}Ir_{3}}$ & 5.70002 & 5.70002 & 2.91013 & $+0.6740$ & 6.817 & -0.127 & 6.690 \\
19 & $\mathrm{EuB_{2}Ni_{3}}$ & 5.19681 & 5.19681 & 2.82962 & $+6.5117$ & 6.862 & -0.138 & 6.724 \\
20 & $\mathrm{EuB_{2}Rh_{3}}$ & 5.64592 & 5.65563 & 2.92039 & $+0.0431$ & 6.640 & -0.085 & 6.555 \\
21 & $\mathrm{EuB_{2}Tc_{3}}$ & 5.72478 & 5.72478 & 3.05958 & $-8.0434$ & 2.146 & -0.184 & 1.962 \\
22 & $\mathrm{EuCd_{2}Ne_{3}}$ & 5.68748 & 5.68748 & 6.08206 & $+0.7744$ & 7.085 & -0.025 & 7.060 \\
23 & $\mathrm{EuCd_{2}Rh_{3}}$ & 5.60256 & 5.60256 & 4.40619 & $-0.2976$ & 6.944 & -0.088 & 6.856 \\
24 & $\mathrm{EuCr_{2}Si_{3}}$ & 5.47785 & 5.47785 & 4.17468 & $+1.3730$ & 14.145 & -0.079 & 14.066 \\
25 & $\mathrm{EuCu_{2}Al_{3}}$ & 5.48067 & 5.48067 & 4.12581 & $-0.3880$ & 6.845 & -0.119 & 6.726 \\
26 & $\mathrm{EuCu_{2}Au_{3}}$ & 5.51735 & 5.51735 & 4.40099 & $+1.2429$ & 6.917 & -0.046 & 6.871 \\
27 & $\mathrm{EuCu_{2}Ga_{3}}$ & 5.46427 & 5.46427 & 4.12586 & $-0.2748$ & 6.959 & -0.087 & 6.872 \\
28 & $\mathrm{EuCu_{2}Ne_{3}}$ & 5.27689 & 5.27689 & 5.41180 & $+0.7029$ & 6.924 & -0.024 & 6.900 \\
29 & $\mathrm{EuCu_{2}Si_{3}}$ & 5.24388 & 5.24388 & 4.05119 & $+1.3111$ & 6.855 & -0.074 & 6.781 \\
30 & $\mathrm{EuCu_{2}Zn_{3}}$ & 5.40535 & 5.40535 & 4.16627 & $+1.0732$ & 6.843 & -0.074 & 6.769 \\
31 & $\mathrm{EuCu_{5}}$ & 5.13964 & 5.13964 & 4.19608 & $+3.1087$ & 7.067 & -0.096 & 6.971 \\
32 & $\mathrm{EuGa_{2}Cu_{3}}$ & 5.52457 & 5.52457 & 3.96572 & $+2.8567$ & 6.923 & -0.065 & 6.858 \\
33 & $\mathrm{EuGa_{2}Ni_{3}}$ & 5.36045 & 5.36046 & 3.72619 & $+2.3258$ & 6.118 & -0.173 & 5.945 \\
34 & $\mathrm{EuGa_{2}Pd_{3}}$ & 5.63805 & 5.63806 & 3.97483 & $+3.6032$ & 7.113 & -0.037 & 7.076 \\
35 & $\mathrm{EuGa_{2}Rh_{3}}$ & 5.67243 & 5.67243 & 3.79647 & $+3.3243$ & 6.488 & -0.273 & 6.215 \\
36 & $\mathrm{EuGe_{2}Ni_{3}}$ & 5.52202 & 5.52196 & 3.52837 & $-0.5275$ & 7.052 & -0.073 & 6.979 \\
37 & $\mathrm{EuGe_{2}Pd_{3}}$ & 5.84631 & 5.84631 & 3.80004 & $+0.4255$ & 7.069 & -0.029 & 7.040 \\
38 & $\mathrm{EuGe_{2}Rh_{3}}$ & 5.74190 & 5.74190 & 3.71373 & $+2.4691$ & 6.842 & -0.266 & 6.576 \\
39 & $\mathrm{EuGe_{2}Tc_{3}}$ & 5.80029 & 5.80089 & 3.95220 & $-2.1938$ & 0.633 & -0.186 & 0.447 \\
40 & $\mathrm{EuHf_{2}Ir_{3}}$ & 5.98233 & 5.98233 & 4.11306 & $+9.0855$ & 7.031 & -0.252 & 6.779 \\
41 & $\mathrm{EuHg_{2}Ne_{3}}$ & 5.72465 & 5.72465 & 6.02308 & $+2.0609$ & 7.162 & -0.048 & 7.114 \\
42 & $\mathrm{EuIn_{2}Rh_{3}}$ & 5.80361 & 5.80361 & 4.17780 & $+3.4645$ & 7.259 & -0.072 & 7.187 \\
43 & $\mathrm{EuIr_{2}Re_{3}}$ & 5.23366 & 5.23366 & 4.53452 & $-1.5275$ & 7.330 & 0.099 & 7.429 \\
44 & $\mathrm{EuIr_{2}Si_{3}}$ & 5.34400 & 5.34400 & 3.97623 & $-1.2638$ & 6.580 & -0.202 & 6.378 \\
45 & $\mathrm{EuIr_{2}Zn_{3}}$ & 5.47056 & 5.47057 & 4.09952 & $+14.0935$ & 5.959 & -0.772 & 5.187 \\
46 & $\mathrm{EuMn_{2}Rh_{3}}$ & 5.57837 & 5.57837 & 4.10641 & $-3.3923$ & 14.732 & -0.308 & 14.424 \\
47 & $\mathrm{EuMo_{2}Rh_{3}}$ & 5.69070 & 5.69070 & 4.23681 & $-3.7424$ & 12.230 & -0.334 & 11.896 \\
48 & $\mathrm{EuMo_{2}Ru_{3}}$ & 5.73486 & 5.73537 & 3.84377 & $+3.2496$ & 6.651 & -0.188 & 6.463 \\
49 & $\mathrm{EuMo_{2}Si_{3}}$ & 5.70051 & 5.70051 & 4.09141 & $-2.8661$ & 1.839 & -0.023 & 1.816 \\
50 & $\mathrm{EuNb_{2}Ir_{3}}$ & 5.64765 & 5.64765 & 4.17826 & $+1.7803$ & 6.268 & -0.291 & 5.977 \\
51 & $\mathrm{EuNb_{2}Pt_{3}}$ & 5.54169 & 5.54169 & 4.43445 & $+2.8349$ & 6.987 & -0.052 & 6.935 \\
52 & $\mathrm{EuNb_{2}Rh_{3}}$ & 5.64761 & 5.64761 & 4.13444 & $+2.8980$ & 6.309 & -0.237 & 6.072 \\
53 & $\mathrm{EuNb_{2}Tc_{3}}$ & 5.88599 & 5.88603 & 4.04039 & $-2.0132$ & 3.398 & -0.237 & 3.161 \\
54 & $\mathrm{EuNe_{2}Ni_{3}}$ & 5.09462 & 5.09461 & 4.99967 & $-0.0417$ & 3.932 & -0.050 & 3.882 \\
55 & $\mathrm{EuNe_{2}Pt_{3}}$ & 5.19284 & 5.19284 & 5.19616 & $+0.4200$ & 6.444 & -0.099 & 6.345 \\
56 & $\mathrm{EuNe_{2}Si_{3}}$ & 5.07745 & 5.07745 & 5.33937 & $+0.1488$ & 6.954 & -0.004 & 6.950 \\
57 & $\mathrm{EuNi_{2}Cu_{3}}$ & 5.04353 & 5.04353 & 3.97941 & $-5.1199$ & 6.848 & -0.105 & 6.743 \\
58 & $\mathrm{EuNi_{2}Ge_{3}}$ & 5.45566 & 5.45566 & 3.84725 & $+1.1502$ & 6.800 & -0.082 & 6.718 \\
59 & $\mathrm{EuNi_{2}In_{3}}$ & 6.17342 & 6.17342 & 3.81344 & $+1.8635$ & 7.509 & -0.006 & 7.503 \\
60 & $\mathrm{EuNi_{2}Si_{3}}$ & 5.18175 & 5.18175 & 3.75069 & $+2.9287$ & 6.805 & -0.115 & 6.690 \\
61 & $\mathrm{EuOs_{2}Al_{3}}$ & 5.61177 & 5.61178 & 3.98285 & $+9.6522$ & 5.631 & -0.501 & 5.130 \\
62 & $\mathrm{EuOs_{2}Tc_{3}}$ & 5.27162 & 5.27162 & 4.41792 & $-1.1475$ & 9.448 & -0.141 & 9.307 \\
63 & $\mathrm{EuP_{2}Cr_{3}}$ & 6.16856 & 6.16856 & 3.69904 & $-3.4507$ & 4.487 & -0.016 & 4.471 \\
64 & $\mathrm{EuP_{2}Hf_{3}}$ & 6.52504 & 6.52503 & 3.84405 & $+0.1778$ & 6.796 & -0.052 & 6.744 \\
65 & $\mathrm{EuP_{2}Mn_{3}}$ & 5.99587 & 5.99587 & 3.77762 & $-3.0359$ & 5.462 & 0.054 & 5.516 \\
66 & $\mathrm{EuP_{2}Pd_{3}}$ & 5.72791 & 5.72791 & 3.67728 & $+1.9006$ & 6.945 & -0.053 & 6.892 \\
67 & $\mathrm{EuP_{2}Re_{3}}$ & 5.48883 & 5.48881 & 3.83876 & $+4.3806$ & 6.615 & -0.158 & 6.457 \\
68 & $\mathrm{EuP_{2}Rh_{3}}$ & 5.65982 & 5.65977 & 3.52895 & $+1.4204$ & 6.067 & -0.134 & 5.933 \\
69 & $\mathrm{EuP_{2}V_{3}}$ & 6.15274 & 6.15274 & 3.62543 & $-0.7836$ & 15.189 & -0.078 & 15.111 \\
70 & $\mathrm{EuP_{2}Zr_{3}}$ & 6.61300 & 6.61300 & 3.83025 & $+0.1046$ & 6.886 & -0.042 & 6.844 \\
71 & $\mathrm{EuPd_{2}Ag_{3}}$ & 5.54001 & 5.54001 & 4.56640 & $-1.0339$ & 6.888 & -0.049 & 6.839 \\
72 & $\mathrm{EuPd_{2}Al_{3}}$ & 5.53776 & 5.53776 & 4.17347 & $-0.2341$ & 6.911 & -0.112 & 6.799 \\
73 & $\mathrm{EuPd_{2}Au_{3}}$ & 5.53668 & 5.53668 & 4.55153 & $-2.9506$ & 6.818 & -0.067 & 6.751 \\
74 & $\mathrm{EuPd_{2}Cu_{3}}$ & 5.24077 & 5.24077 & 4.23071 & $-1.8633$ & 6.889 & -0.079 & 6.810 \\
75 & $\mathrm{EuPd_{2}Pt_{3}}$ & 5.38609 & 5.38728 & 4.37405 & $+10.3864$ & 6.438 & -0.433 & 6.005 \\
76 & $\mathrm{EuPd_{2}Si_{3}}$ & 5.34430 & 5.34430 & 4.09308 & $+1.6309$ & 6.823 & -0.088 & 6.735 \\
77 & $\mathrm{EuPt_{2}Cu_{3}}$ & 5.30365 & 5.30365 & 4.18453 & $+7.8214$ & 6.694 & -0.626 & 6.068 \\
78 & $\mathrm{EuPt_{2}Pd_{3}}$ & 5.38047 & 5.38047 & 4.37725 & $+20.2282$ & 6.532 & -0.412 & 6.120 \\
79 & $\mathrm{EuPt_{2}Ta_{3}}$ & 6.01465 & 6.01465 & 4.26217 & $+10.4404$ & 9.553 & -0.436 & 9.117 \\
80 & $\mathrm{EuPt_{2}Zn_{3}}$ & 5.55052 & 5.55052 & 4.13558 & $+8.0932$ & 6.759 & -0.178 & 6.581 \\
81 & $\mathrm{EuRe_{5}}$ & 5.33015 & 5.33015 & 4.43496 & $-6.1603$ & 6.850 & -0.125 & 6.725 \\
82 & $\mathrm{EuRh_{2}Al_{3}}$ & 5.67544 & 5.67544 & 3.90456 & $-0.2969$ & 6.964 & -0.115 & 6.849 \\
83 & $\mathrm{EuRh_{2}Ge_{3}}$ & 5.61688 & 5.61688 & 4.00440 & $-2.0119$ & 6.803 & -0.118 & 6.685 \\
84 & $\mathrm{EuRh_{2}Ir_{3}}$ & 5.29538 & 5.29538 & 4.29722 & $-2.5005$ & 6.200 & -0.361 & 5.839 \\
85 & $\mathrm{EuRh_{2}Pd_{3}}$ & 5.30224 & 5.30224 & 4.37767 & $+10.2295$ & 4.276 & -0.559 & 3.717 \\
86 & $\mathrm{EuRh_{2}Pt_{3}}$ & 5.36536 & 5.36536 & 4.34734 & $+17.9853$ & 5.959 & -0.418 & 5.541 \\
87 & $\mathrm{EuRh_{2}Si_{3}}$ & 5.29739 & 5.29739 & 3.99746 & $-4.0471$ & 6.791 & -0.140 & 6.651 \\
88 & $\mathrm{EuRh_{2}Ti_{3}}$ & 6.55501 & 6.55501 & 3.74179 & $-0.1796$ & 11.452 & -0.120 & 11.332 \\
89 & $\mathrm{EuRh_{2}Zn_{3}}$ & 5.49324 & 5.49324 & 4.06888 & $+7.1460$ & 6.688 & -0.404 & 6.284 \\
90 & $\mathrm{EuRu_{2}Al_{3}}$ & 5.60173 & 5.60122 & 3.91872 & $+6.3713$ & 6.735 & -0.296 & 6.439 \\
91 & $\mathrm{EuRu_{2}Nb_{3}}$ & 6.15231 & 6.15231 & 3.77601 & $-2.8779$ & 6.812 & -0.136 & 6.676 \\
92 & $\mathrm{EuRu_{2}Si_{3}}$ & 5.35644 & 5.35644 & 3.88126 & $-4.5886$ & 7.509 & -0.033 & 7.476 \\
93 & $\mathrm{EuRu_{2}Zn_{3}}$ & 5.59322 & 5.59331 & 4.04080 & $-5.7287$ & 3.922 & -0.261 & 3.661 \\
94 & $\mathrm{EuSi_{2}Al_{3}}$ & 5.67377 & 5.67377 & 4.11292 & $+3.4930$ & 6.797 & -0.113 & 6.684 \\
95 & $\mathrm{EuSi_{2}Au_{3}}$ & 6.04180 & 6.04180 & 3.80772 & $+2.6052$ & 6.855 & -0.059 & 6.796 \\
96 & $\mathrm{EuSi_{2}Cu_{3}}$ & 5.41929 & 5.41929 & 3.83165 & $+6.5391$ & 6.945 & -0.150 & 6.795 \\
97 & $\mathrm{EuSi_{2}Ir_{3}}$ & 5.65700 & 5.65701 & 3.61412 & $+12.0510$ & 6.694 & -0.362 & 6.332 \\
98 & $\mathrm{EuSi_{2}Ni_{3}}$ & 5.39402 & 5.39379 & 3.41295 & $+1.8673$ & 7.022 & -0.098 & 6.924 \\
99 & $\mathrm{EuSi_{2}Pd_{3}}$ & 5.75202 & 5.75202 & 3.66423 & $+1.0261$ & 6.962 & -0.054 & 6.908 \\
100 & $\mathrm{EuSi_{2}Pt_{3}}$ & 5.88240 & 5.88240 & 3.59833 & $+0.4923$ & 7.015 & -0.011 & 7.004 \\
101 & $\mathrm{EuSi_{2}Rh_{3}}$ & 5.66769 & 5.66768 & 3.57261 & $+11.0454$ & 6.759 & -0.284 & 6.475 \\
102 & $\mathrm{EuSi_{2}Zn_{3}}$ & 5.54525 & 5.54525 & 4.12263 & $+1.2517$ & 6.846 & -0.085 & 6.761 \\
103 & $\mathrm{EuSn_{2}Ni_{3}}$ & 5.76680 & 5.76680 & 3.84983 & $-2.0775$ & 8.003 & 0.023 & 8.026 \\
104 & $\mathrm{EuTa_{2}Ir_{3}}$ & 5.71569 & 5.71569 & 4.17667 & $+3.4265$ & 6.113 & -0.238 & 5.875 \\
105 & $\mathrm{EuTa_{2}Pt_{3}}$ & 5.70282 & 5.70282 & 4.33630 & $-1.1982$ & 7.048 & -0.098 & 6.950 \\
106 & $\mathrm{EuTa_{2}Rh_{3}}$ & 5.72088 & 5.72088 & 4.13433 & $+13.2471$ & 6.379 & -0.208 & 6.171 \\
107 & $\mathrm{EuTc_{2}Al_{3}}$ & 5.60428 & 5.60428 & 4.03558 & $+0.9114$ & 8.736 & -0.167 & 8.569 \\
108 & $\mathrm{EuTc_{2}B_{3}}$ & 5.51065 & 5.51065 & 2.97145 & $+1.5568$ & 6.299 & -0.136 & 6.163 \\
109 & $\mathrm{EuTc_{2}Ge_{3}}$ & 5.74625 & 5.74625 & 4.01223 & $-5.7189$ & 3.393 & -0.142 & 3.251 \\
110 & $\mathrm{EuTc_{2}Mo_{3}}$ & 6.19731 & 6.19731 & 4.07930 & $-0.0713$ & 10.584 & -0.232 & 10.352 \\
111 & $\mathrm{EuTc_{2}Re_{3}}$ & 5.30803 & 5.30803 & 4.41554 & $+12.1310$ & 7.685 & -0.280 & 7.405 \\
112 & $\mathrm{EuTc_{5}}$ & 5.76195 & 5.76208 & 3.93609 & $-1.9660$ & 5.914 & -0.209 & 5.705 \\
113 & $\mathrm{EuTi_{2}Rh_{3}}$ & 5.88124 & 5.88124 & 3.82761 & $+1.3978$ & 7.002 & -0.181 & 6.821 \\
114 & $\mathrm{EuTl_{2}Pd_{3}}$ & 5.90268 & 5.90268 & 4.37660 & $-0.4860$ & 7.059 & -0.026 & 7.033 \\
115 & $\mathrm{EuZn_{2}Cu_{3}}$ & 5.43888 & 5.43888 & 4.01689 & $+0.4912$ & 7.065 & -0.062 & 7.003 \\
116 & $\mathrm{EuZn_{2}Ni_{3}}$ & 5.21735 & 5.21735 & 3.81012 & $+7.3254$ & 6.841 & -0.149 & 6.692 \\
117 & $\mathrm{EuZn_{2}Pd_{3}}$ & 5.60873 & 5.60873 & 3.96761 & $+1.3712$ & 6.877 & -0.060 & 6.817 \\
118 & $\mathrm{EuZn_{2}Pt_{3}}$ & 5.57728 & 5.57728 & 4.04430 & $+0.5005$ & 6.820 & -0.100 & 6.720 \\
119 & $\mathrm{EuZn_{2}Rh_{3}}$ & 5.49659 & 5.49660 & 4.00350 & $+0.7276$ & 6.528 & -0.231 & 6.297 \\
120 & $\mathrm{EuZn_{5}}$ & 5.38317 & 5.38317 & 4.41155 & $+0.9765$ & 6.943 & -0.058 & 6.885 \\
121 & $\mathrm{EuZr_{2}Rh_{3}}$ & 6.14322 & 6.14322 & 3.99163 & $+3.1080$ & 6.658 & -0.089 & 6.569 \\
122 & $\mathrm{GdAg_{2}Cd_{3}}$ & 5.73174 & 5.73174 & 4.78150 & $-0.0388$ & 7.163 & -0.012 & 7.151 \\
123 & $\mathrm{GdAg_{2}Zn_{3}}$ & 5.35861 & 5.35861 & 4.52255 & $-0.0586$ & 7.182 & 0.000 & 7.182 \\
124 & $\mathrm{GdAg_{5}}$ & 5.58431 & 5.58431 & 4.66875 & $+0.9379$ & 7.444 & -0.013 & 7.431 \\
125 & $\mathrm{GdAl_{2}Ag_{3}}$ & 5.62185 & 5.62185 & 4.32557 & $+2.6025$ & 7.499 & -0.008 & 7.491 \\
126 & $\mathrm{GdAl_{2}Cu_{3}}$ & 5.30781 & 5.30781 & 4.03198 & $-0.4682$ & 7.472 & -0.009 & 7.463 \\
127 & $\mathrm{GdAl_{2}Fe_{3}}$ & 5.45125 & 5.45124 & 4.21478 & $-2.6507$ & 1.762 & 0.075 & 1.837 \\
128 & $\mathrm{GdAl_{2}Ir_{3}}$ & 5.56742 & 5.56742 & 3.80868 & $+0.4186$ & 6.816 & -0.053 & 6.763 \\
129 & $\mathrm{GdAl_{2}Pd_{3}}$ & 5.59989 & 5.59989 & 3.88221 & $-1.2593$ & 7.008 & -0.001 & 7.007 \\
130 & $\mathrm{GdAl_{2}Re_{3}}$ & 5.33552 & 5.33552 & 4.32529 & $+0.7930$ & 6.873 & -0.018 & 6.855 \\
131 & $\mathrm{GdAl_{2}Tc_{3}}$ & 5.34833 & 5.34832 & 4.27490 & $+1.7783$ & 5.051 & 0.038 & 5.089 \\
132 & $\mathrm{GdAl_{2}Zn_{3}}$ & 5.40521 & 5.40521 & 4.24182 & $+0.3793$ & 7.013 & -0.004 & 7.009 \\
133 & $\mathrm{GdAl_{5}}$ & 5.54339 & 5.54339 & 4.39467 & $+0.4514$ & 7.286 & -0.001 & 7.285 \\
134 & $\mathrm{GdAs_{2}Ni_{3}}$ & 5.41304 & 5.41303 & 3.53568 & $-0.6302$ & 7.129 & 0.012 & 7.141 \\
135 & $\mathrm{GdAs_{2}Pd_{3}}$ & 5.67831 & 5.67831 & 3.91084 & $-0.1871$ & 7.296 & 0.010 & 7.306 \\
136 & $\mathrm{GdAu_{2}Cu_{3}}$ & 5.29104 & 5.29104 & 4.32938 & $+2.2728$ & 7.045 & -0.016 & 7.029 \\
137 & $\mathrm{GdAu_{2}Zn_{3}}$ & 5.35719 & 5.35719 & 4.49695 & $+0.8552$ & 7.297 & -0.003 & 7.294 \\
138 & $\mathrm{GdAu_{5}}$ & 5.52289 & 5.52289 & 4.73670 & $-0.8148$ & 7.381 & 0.018 & 7.399 \\
139 & $\mathrm{GdB_{2}Co_{3}}$ & 5.05328 & 5.05330 & 2.94551 & $-0.6005$ & 5.655 & -0.004 & 5.651 \\
140 & $\mathrm{GdB_{2}Fe_{3}}$ & 5.55253 & 5.55253 & 3.03207 & $-0.7222$ & 1.624 & 0.084 & 1.708 \\
141 & $\mathrm{GdB_{2}Ir_{3}}$ & 5.63112 & 5.63112 & 2.93504 & $+0.6311$ & 7.124 & -0.012 & 7.112 \\
142 & $\mathrm{GdB_{2}Mn_{3}}$ & 5.49243 & 5.49242 & 3.29872 & $-2.4303$ & 17.502 & 0.015 & 17.517 \\
143 & $\mathrm{GdB_{2}Ni_{3}}$ & 5.12534 & 5.12534 & 2.84498 & $+0.0121$ & 7.236 & 0.008 & 7.244 \\
144 & $\mathrm{GdB_{2}Re_{3}}$ & 5.52532 & 5.52532 & 3.19297 & $-0.4042$ & 7.078 & 0.024 & 7.102 \\
145 & $\mathrm{GdB_{2}Ru_{3}}$ & 5.47965 & 5.47965 & 2.98973 & $-0.3099$ & 6.069 & -0.094 & 5.975 \\
146 & $\mathrm{GdCd_{2}Ag_{3}}$ & 5.77672 & 5.77672 & 4.64796 & $-0.2364$ & 7.090 & -0.001 & 7.089 \\
147 & $\mathrm{GdCd_{2}Ir_{3}}$ & 5.49596 & 5.49596 & 4.43975 & $+1.9235$ & 7.262 & 0.150 & 7.412 \\
148 & $\mathrm{GdCd_{2}Ne_{3}}$ & 5.45622 & 5.45622 & 5.89127 & $-0.9084$ & 7.440 & -0.028 & 7.412 \\
149 & $\mathrm{GdCo_{2}Cu_{3}}$ & 4.96477 & 4.96477 & 4.20431 & $-0.8788$ & 10.490 & 0.084 & 10.574 \\
150 & $\mathrm{GdCo_{2}Si_{3}}$ & 5.07832 & 5.07833 & 3.73415 & $-0.0897$ & 6.296 & -0.043 & 6.253 \\
151 & $\mathrm{GdCu_{2}Al_{3}}$ & 5.41316 & 5.41316 & 4.02939 & $-0.1400$ & 7.163 & 0.000 & 7.163 \\
152 & $\mathrm{GdCu_{2}Ga_{3}}$ & 5.40316 & 5.40315 & 4.02343 & $-0.1463$ & 7.058 & 0.002 & 7.060 \\
153 & $\mathrm{GdCu_{2}Ni_{3}}$ & 4.94965 & 4.94976 & 3.87930 & $-0.5859$ & 4.864 & -0.039 & 4.825 \\
154 & $\mathrm{GdCu_{2}Zn_{3}}$ & 5.18352 & 5.18353 & 4.23323 & $+0.5672$ & 7.172 & -0.011 & 7.161 \\
155 & $\mathrm{GdCu_{5}}$ & 5.08511 & 5.08511 & 4.07860 & $+0.8560$ & 7.470 & 0.005 & 7.475 \\
156 & $\mathrm{GdFe_{2}Co_{3}}$ & 5.18806 & 5.18799 & 3.89638 & $-1.7785$ & 4.204 & 0.353 & 4.557 \\
157 & $\mathrm{GdGa_{2}Ir_{3}}$ & 5.56808 & 5.56808 & 3.86856 & $-1.3100$ & 6.817 & -0.049 & 6.768 \\
158 & $\mathrm{GdGa_{2}Ni_{3}}$ & 5.29868 & 5.29869 & 3.67865 & $+0.0275$ & 6.481 & -0.073 & 6.408 \\
159 & $\mathrm{GdGa_{2}Pd_{3}}$ & 5.60965 & 5.60965 & 3.90241 & $+0.1875$ & 7.091 & -0.004 & 7.087 \\
160 & $\mathrm{GdGa_{2}Rh_{3}}$ & 5.60210 & 5.60210 & 3.77910 & $+0.3336$ & 6.836 & -0.030 & 6.806 \\
161 & $\mathrm{GdGa_{2}Zn_{3}}$ & 5.56975 & 5.56975 & 4.11694 & $-0.1945$ & 7.111 & 0.002 & 7.113 \\
162 & $\mathrm{GdGa_{5}}$ & 5.60578 & 5.60578 & 4.52849 & $-1.2288$ & 7.367 & -0.006 & 7.361 \\
163 & $\mathrm{GdGe_{2}Al_{3}}$ & 5.66541 & 5.66541 & 4.28588 & $-0.8128$ & 7.083 & 0.016 & 7.099 \\
164 & $\mathrm{GdGe_{2}Cu_{3}}$ & 5.42941 & 5.42941 & 3.92089 & $+1.0033$ & 7.178 & 0.017 & 7.195 \\
165 & $\mathrm{GdGe_{2}Fe_{3}}$ & 5.57560 & 5.40405 & 4.14792 & $-1.9046$ & 2.336 & 0.090 & 2.426 \\
166 & $\mathrm{GdGe_{2}Ir_{3}}$ & 5.61114 & 5.61114 & 3.81961 & $+0.8309$ & 6.908 & -0.018 & 6.890 \\
167 & $\mathrm{GdGe_{2}Ni_{3}}$ & 5.41710 & 5.41710 & 3.58338 & $-0.4137$ & 8.911 & 0.126 & 9.037 \\
168 & $\mathrm{GdGe_{2}Pd_{3}}$ & 5.68196 & 5.68197 & 3.84825 & $-1.7469$ & 8.078 & 0.033 & 8.111 \\
169 & $\mathrm{GdGe_{2}Rh_{3}}$ & 5.59011 & 5.59011 & 3.79598 & $+1.1368$ & 6.870 & -0.007 & 6.863 \\
170 & $\mathrm{GdGe_{2}Zn_{3}}$ & 5.45743 & 5.45743 & 4.39565 & $+0.0746$ & 7.035 & 0.009 & 7.044 \\
171 & $\mathrm{GdIn_{2}Ir_{3}}$ & 5.70874 & 5.70874 & 4.21327 & $-0.9474$ & 6.689 & -0.116 & 6.573 \\
172 & $\mathrm{GdIn_{2}Pd_{3}}$ & 5.81624 & 5.81624 & 4.19919 & $-0.7411$ & 7.128 & 0.003 & 7.131 \\
173 & $\mathrm{GdIr_{2}Al_{3}}$ & 5.41076 & 5.41076 & 4.10446 & $+0.1645$ & 7.118 & 0.024 & 7.142 \\
174 & $\mathrm{GdIr_{2}Fe_{3}}$ & 5.35399 & 5.35399 & 4.24880 & $+10.1843$ & 4.209 & 0.434 & 4.643 \\
175 & $\mathrm{GdIr_{2}Nb_{3}}$ & 5.65017 & 5.65014 & 4.34717 & $-4.1073$ & 6.855 & -0.041 & 6.814 \\
176 & $\mathrm{GdIr_{2}Re_{3}}$ & 5.21224 & 5.21224 & 4.51413 & $+3.5937$ & 7.047 & 0.028 & 7.075 \\
177 & $\mathrm{GdIr_{2}Si_{3}}$ & 5.28596 & 5.28596 & 3.97031 & $+3.1346$ & 7.255 & 0.053 & 7.308 \\
178 & $\mathrm{GdIr_{2}Ta_{3}}$ & 5.73528 & 5.73528 & 4.36357 & $-2.9404$ & 9.134 & -0.355 & 8.779 \\
179 & $\mathrm{GdIr_{2}Tc_{3}}$ & 5.20410 & 5.20393 & 4.51282 & $+26.2708$ & 2.312 & -0.653 & 1.659 \\
180 & $\mathrm{GdIr_{2}Zn_{3}}$ & 5.34953 & 5.34953 & 4.14015 & $+4.5837$ & 7.322 & 0.182 & 7.504 \\
181 & $\mathrm{GdIr_{5}}$ & 5.27151 & 5.27151 & 4.32293 & $+8.7696$ & 7.189 & 0.184 & 7.373 \\
182 & $\mathrm{GdMo_{2}Si_{3}}$ & 5.39480 & 5.39480 & 4.06987 & $-0.1961$ & 6.567 & 0.047 & 6.614 \\
183 & $\mathrm{GdNb_{2}Ir_{3}}$ & 5.56974 & 5.56974 & 4.21098 & $+2.7820$ & 7.022 & -0.075 & 6.947 \\
184 & $\mathrm{GdNb_{2}Mo_{3}}$ & 6.01522 & 6.01718 & 4.18295 & $-0.7315$ & 1.637 & 0.045 & 1.682 \\
185 & $\mathrm{GdNb_{2}Re_{3}}$ & 5.44924 & 5.44922 & 4.50369 & $+0.4401$ & 6.883 & 0.025 & 6.908 \\
186 & $\mathrm{GdNb_{2}Tc_{3}}$ & 5.77088 & 5.77088 & 4.04054 & $-0.3108$ & 4.585 & -0.116 & 4.469 \\
187 & $\mathrm{GdNe_{2}Si_{3}}$ & 5.03369 & 5.03369 & 5.11726 & $+0.2868$ & 7.179 & 0.016 & 7.195 \\
188 & $\mathrm{GdNi_{2}Al_{3}}$ & 5.40841 & 5.40841 & 3.76693 & $+0.6052$ & 7.303 & 0.014 & 7.317 \\
189 & $\mathrm{GdNi_{2}Fe_{3}}$ & 5.11366 & 5.11366 & 4.23698 & $-0.0925$ & 3.268 & 0.144 & 3.412 \\
190 & $\mathrm{GdNi_{2}Ga_{3}}$ & 5.34454 & 5.34454 & 3.82598 & $+0.8002$ & 7.779 & 0.033 & 7.812 \\
191 & $\mathrm{GdNi_{2}Ge_{3}}$ & 5.40777 & 5.40782 & 3.80584 & $-0.1254$ & 7.024 & 0.005 & 7.029 \\
192 & $\mathrm{GdNi_{2}Si_{3}}$ & 5.13904 & 5.13904 & 3.70113 & $-0.2336$ & 7.316 & 0.011 & 7.327 \\
193 & $\mathrm{GdNi_{2}Tc_{3}}$ & 5.03283 & 5.03283 & 4.35630 & $+2.4490$ & 3.627 & -0.087 & 3.540 \\
194 & $\mathrm{GdNi_{2}Ti_{3}}$ & 6.48307 & 6.48307 & 3.49267 & $-0.6903$ & 2.653 & 0.070 & 2.723 \\
195 & $\mathrm{GdNi_{2}Zn_{3}}$ & 5.19099 & 5.19099 & 3.94610 & $+1.1464$ & 7.260 & 0.016 & 7.276 \\
196 & $\mathrm{GdNi_{5}}$ & 4.86534 & 4.86537 & 4.12329 & $+1.8455$ & 4.454 & -0.050 & 4.404 \\
197 & $\mathrm{GdOs_{2}Nb_{3}}$ & 5.76506 & 5.76506 & 4.19770 & $-1.1078$ & 6.781 & -0.036 & 6.745 \\
198 & $\mathrm{GdOs_{2}Re_{3}}$ & 5.23569 & 5.23305 & 4.51041 & $-15.0407$ & 4.342 & -0.393 & 3.949 \\
199 & $\mathrm{GdOs_{2}Si_{3}}$ & 5.23340 & 5.23340 & 4.03888 & $+2.8428$ & 6.324 & -0.137 & 6.187 \\
200 & $\mathrm{GdOs_{2}Ta_{3}}$ & 5.81367 & 5.81367 & 4.25840 & $+0.7646$ & 6.725 & -0.020 & 6.705 \\
201 & $\mathrm{GdP_{2}Cu_{3}}$ & 5.37518 & 5.37518 & 3.69335 & $+0.0269$ & 7.153 & 0.008 & 7.161 \\
202 & $\mathrm{GdP_{2}Ir_{3}}$ & 5.58291 & 5.58291 & 3.58424 & $-0.1608$ & 7.095 & -0.009 & 7.086 \\
203 & $\mathrm{GdP_{2}Rh_{3}}$ & 5.55657 & 5.55656 & 3.56365 & $-2.1785$ & 7.935 & 0.062 & 7.997 \\
204 & $\mathrm{GdP_{2}Ru_{3}}$ & 5.53982 & 5.53981 & 3.47016 & $-0.1422$ & 6.448 & -0.027 & 6.421 \\
205 & $\mathrm{GdP_{2}Tc_{3}}$ & 5.65924 & 5.65805 & 3.58843 & $+4.5089$ & 2.882 & -0.158 & 2.724 \\
206 & $\mathrm{GdPd_{2}Al_{3}}$ & 5.39806 & 5.39806 & 4.19274 & $+0.1766$ & 7.307 & 0.004 & 7.311 \\
207 & $\mathrm{GdPd_{2}Au_{3}}$ & 5.40178 & 5.40178 & 4.61485 & $+0.4581$ & 7.137 & 0.006 & 7.143 \\
208 & $\mathrm{GdPd_{2}Ga_{3}}$ & 5.38041 & 5.38041 & 4.22082 & $-0.0845$ & 7.594 & 0.010 & 7.604 \\
209 & $\mathrm{GdPd_{2}Pt_{3}}$ & 5.31083 & 5.31083 & 4.39850 & $+0.0044$ & 6.850 & -0.022 & 6.828 \\
210 & $\mathrm{GdPd_{2}Si_{3}}$ & 5.28774 & 5.28774 & 4.09310 & $+0.0403$ & 7.150 & 0.010 & 7.160 \\
211 & $\mathrm{GdPd_{2}Zn_{3}}$ & 5.33335 & 5.33335 & 4.21574 & $-1.1586$ & 7.159 & 0.000 & 7.159 \\
212 & $\mathrm{GdPt_{2}Cu_{3}}$ & 5.13243 & 5.13243 & 4.31329 & $+7.2615$ & 7.081 & 0.039 & 7.120 \\
213 & $\mathrm{GdPt_{5}}$ & 5.35073 & 5.35073 & 4.42199 & $-0.8133$ & 6.730 & -0.064 & 6.666 \\
214 & $\mathrm{GdRe_{2}Al_{3}}$ & 5.42649 & 5.42649 & 4.19046 & $+6.6416$ & 7.108 & -0.026 & 7.082 \\
215 & $\mathrm{GdRe_{2}Ir_{3}}$ & 5.27961 & 5.27963 & 4.35201 & $+3.6200$ & 7.320 & -0.082 & 7.238 \\
216 & $\mathrm{GdRe_{5}}$ & 5.26246 & 5.26247 & 4.49237 & $+0.8936$ & 6.996 & -0.029 & 6.967 \\
217 & $\mathrm{GdRh_{2}Al_{3}}$ & 5.45542 & 5.45542 & 4.04457 & $+0.0569$ & 7.476 & 0.018 & 7.494 \\
218 & $\mathrm{GdRh_{2}Cu_{3}}$ & 5.00618 & 5.00618 & 4.34056 & $+3.3473$ & 7.574 & 0.142 & 7.716 \\
219 & $\mathrm{GdRh_{2}Ga_{3}}$ & 5.46036 & 5.46036 & 4.05635 & $-2.4482$ & 7.679 & 0.033 & 7.712 \\
220 & $\mathrm{GdRh_{2}Ir_{3}}$ & 5.23409 & 5.23409 & 4.32236 & $+0.9749$ & 7.232 & 0.070 & 7.302 \\
221 & $\mathrm{GdRh_{2}Si_{3}}$ & 5.23685 & 5.23685 & 3.97964 & $+0.6989$ & 6.997 & 0.007 & 7.004 \\
222 & $\mathrm{GdRh_{2}Zn_{3}}$ & 5.33315 & 5.33315 & 4.13516 & $+0.9824$ & 7.198 & 0.017 & 7.215 \\
223 & $\mathrm{GdRu_{2}Al_{3}}$ & 5.49591 & 5.49591 & 3.95438 & $+1.2262$ & 8.246 & 0.107 & 8.353 \\
224 & $\mathrm{GdRu_{2}Mn_{3}}$ & 5.53892 & 5.53891 & 4.26973 & $+2.1646$ & 7.465 & 0.223 & 7.688 \\
225 & $\mathrm{GdRu_{2}Ta_{3}}$ & 6.02208 & 6.02194 & 3.93739 & $-0.0686$ & 6.583 & 0.004 & 6.587 \\
226 & $\mathrm{GdRu_{2}Tc_{3}}$ & 5.17324 & 5.17354 & 4.45590 & $-0.7417$ & 2.425 & -0.185 & 2.240 \\
227 & $\mathrm{GdSi_{2}Cu_{3}}$ & 5.30066 & 5.30066 & 3.78910 & $-0.1705$ & 7.155 & 0.005 & 7.160 \\
228 & $\mathrm{GdSi_{2}Fe_{3}}$ & 5.36338 & 5.36338 & 4.04629 & $-1.3017$ & 16.012 & 0.090 & 16.102 \\
229 & $\mathrm{GdSi_{2}Nb_{3}}$ & 5.61995 & 5.61995 & 4.29435 & $-0.7525$ & 6.980 & 0.000 & 6.980 \\
230 & $\mathrm{GdSi_{2}Ni_{3}}$ & 5.21706 & 5.21706 & 3.49108 & $-0.5799$ & 8.625 & 0.096 & 8.721 \\
231 & $\mathrm{GdSi_{2}Pd_{3}}$ & 5.66366 & 5.66366 & 3.64591 & $+0.1556$ & 7.977 & 0.014 & 7.991 \\
232 & $\mathrm{GdSi_{2}Re_{3}}$ & 5.41547 & 5.41547 & 3.99857 & $+1.6708$ & 6.521 & -0.096 & 6.425 \\
233 & $\mathrm{GdSi_{2}Rh_{3}}$ & 5.51657 & 5.51657 & 3.65983 & $+0.3211$ & 6.904 & -0.003 & 6.901 \\
234 & $\mathrm{GdSi_{2}Ru_{3}}$ & 5.52410 & 5.52410 & 3.61301 & $+1.9936$ & 4.937 & -0.171 & 4.766 \\
235 & $\mathrm{GdSi_{2}Tc_{3}}$ & 5.49029 & 5.49034 & 3.93947 & $-2.5563$ & 2.763 & -0.080 & 2.683 \\
236 & $\mathrm{GdSi_{2}Zn_{3}}$ & 5.37593 & 5.37593 & 4.19324 & $+0.3199$ & 7.164 & -0.014 & 7.150 \\
237 & $\mathrm{GdSn_{2}Cu_{3}}$ & 5.71243 & 5.71243 & 4.22504 & $+0.0999$ & 7.530 & -0.028 & 7.502 \\
238 & $\mathrm{GdTa_{2}Pt_{3}}$ & 5.54651 & 5.54651 & 4.44929 & $-1.1026$ & 7.247 & -0.006 & 7.241 \\
239 & $\mathrm{GdTa_{2}Re_{3}}$ & 5.50888 & 5.50888 & 4.49998 & $-0.5468$ & 6.950 & 0.024 & 6.974 \\
240 & $\mathrm{GdTa_{2}Rh_{3}}$ & 5.62960 & 5.62960 & 4.16793 & $+0.0041$ & 7.085 & -0.048 & 7.037 \\
241 & $\mathrm{GdTa_{2}Ru_{3}}$ & 5.82144 & 5.82144 & 3.87413 & $-0.7948$ & 6.811 & -0.016 & 6.795 \\
242 & $\mathrm{GdTc_{2}Al_{3}}$ & 5.46126 & 5.45108 & 4.11323 & $-0.5878$ & 6.736 & -0.031 & 6.705 \\
243 & $\mathrm{GdTc_{2}Co_{3}}$ & 5.17672 & 5.17673 & 4.14456 & $+0.3052$ & 1.332 & 0.247 & 1.579 \\
244 & $\mathrm{GdTc_{2}Fe_{3}}$ & 5.17754 & 5.17754 & 4.52266 & $+1.8502$ & 1.390 & 0.126 & 1.516 \\
245 & $\mathrm{GdTc_{2}Ge_{3}}$ & 5.55960 & 5.55960 & 4.06489 & $-0.2307$ & 3.686 & -0.140 & 3.546 \\
246 & $\mathrm{GdTc_{2}Mn_{3}}$ & 5.59882 & 5.59882 & 4.45351 & $+1.0084$ & 0.897 & -0.099 & 0.798 \\
247 & $\mathrm{GdTc_{2}Mo_{3}}$ & 5.33587 & 5.33587 & 4.49975 & $+0.2168$ & 7.372 & -0.046 & 7.326 \\
248 & $\mathrm{GdTc_{2}Si_{3}}$ & 5.35999 & 5.35999 & 3.94564 & $-0.3117$ & 8.990 & 0.129 & 9.119 \\
249 & $\mathrm{GdTc_{5}}$ & 5.26035 & 5.26035 & 4.31265 & $-4.9000$ & 7.756 & -0.053 & 7.703 \\
250 & $\mathrm{GdTi_{2}Pd_{3}}$ & 5.74709 & 5.74709 & 4.30346 & $+1.2971$ & 3.728 & -0.008 & 3.720 \\
251 & $\mathrm{GdTi_{2}Rh_{3}}$ & 5.71114 & 5.71322 & 3.95971 & $+0.8438$ & 5.486 & 0.059 & 5.545 \\
252 & $\mathrm{GdTi_{2}Tc_{3}}$ & 5.71966 & 5.71965 & 4.24546 & $-1.0640$ & 4.715 & -0.105 & 4.610 \\
253 & $\mathrm{GdV_{2}Ir_{3}}$ & 5.48863 & 5.48859 & 3.99913 & $-2.8196$ & 5.230 & 0.057 & 5.287 \\
254 & $\mathrm{GdZn_{2}Au_{3}}$ & 5.74916 & 5.74916 & 4.11543 & $+2.8269$ & 7.181 & -0.007 & 7.174 \\
255 & $\mathrm{GdZn_{2}Ir_{3}}$ & 5.39415 & 5.39415 & 4.05834 & $-3.6404$ & 6.869 & -0.048 & 6.821 \\
256 & $\mathrm{GdZn_{2}Pd_{3}}$ & 5.49499 & 5.49499 & 3.97571 & $+0.2009$ & 7.201 & -0.010 & 7.191 \\
257 & $\mathrm{GdZn_{2}Rh_{3}}$ & 5.39964 & 5.39964 & 4.01270 & $-0.6082$ & 6.992 & -0.004 & 6.988 \\
258 & $\mathrm{GdZn_{2}Ru_{3}}$ & 5.25694 & 5.25690 & 4.26489 & $-4.2220$ & 3.022 & -0.170 & 2.852 \\
259 & $\mathrm{GdZn_{5}}$ & 5.31913 & 5.31913 & 4.23623 & $+0.4039$ & 7.051 & -0.003 & 7.048 \\
260 & $\mathrm{LiC_{2}Mn_{3}}$ & 5.54084 & 5.54083 & 2.89265 & $-0.5473$ & 11.263 & 0.005 & 11.268 \\
261 & $\mathrm{LuCo_{5}}$ & 4.93250 & 4.93270 & 3.92644 & $+2.5161$ & 8.569 & 0.424 & 8.993 \\
262 & $\mathrm{LuFe_{2}Co_{3}}$ & 5.07715 & 5.07720 & 3.89566 & $-1.7772$ & 11.073 & 0.354 & 11.427 \\
263 & $\mathrm{LuMn_{2}Fe_{3}}$ & 5.46384 & 5.46384 & 4.40265 & $+0.6044$ & 18.077 & 0.156 & 18.233 \\
264 & $\mathrm{LuNi_{2}Fe_{3}}$ & 5.03439 & 5.03415 & 4.20204 & $+0.4527$ & 10.157 & 0.147 & 10.304 \\
265 & $\mathrm{MgP_{2}Mn_{3}}$ & 6.40093 & 6.40093 & 3.52342 & $-0.1049$ & 13.302 & 0.007 & 13.309 \\
266 & $\mathrm{NdCo_{2}Fe_{3}}$ & 5.35225 & 5.30000 & 4.08912 & $+1.6827$ & 16.115 & -0.179 & 15.936 \\
267 & $\mathrm{NdMn_{2}Co_{3}}$ & 5.51471 & 5.48547 & 4.04991 & $-2.3121$ & 6.347 & -0.618 & 5.729 \\
268 & $\mathrm{NdTc_{2}Fe_{3}}$ & 5.40255 & 5.40839 & 4.48681 & $-7.1212$ & 8.939 & -0.391 & 8.548 \\
269 & $\mathrm{PmGe_{2}Mn_{3}}$ & 5.91705 & 5.91544 & 4.11233 & $-1.5151$ & 7.993 & 0.486 & 8.479 \\
270 & $\mathrm{PmMn_{2}Co_{3}}$ & 5.50700 & 5.50637 & 4.05543 & $-0.1658$ & 9.473 & 0.685 & 10.158 \\
271 & $\mathrm{PmNi_{2}Fe_{3}}$ & 5.20830 & 5.22064 & 4.18108 & $+4.0727$ & 6.274 & 0.603 & 6.877 \\
272 & $\mathrm{PmRh_{2}Mn_{3}}$ & 5.70425 & 5.69981 & 4.17636 & $-2.1545$ & 9.304 & 0.505 & 9.809 \\
273 & $\mathrm{PrCo_{5}}$ & 5.03897 & 5.61392 & 3.89735 & $-1.4051$ & 10.962 & 0.151 & 11.113 \\
274 & $\mathrm{PrRu_{2}Fe_{3}}$ & 5.37259 & 5.37259 & 4.25179 & $-6.2080$ & 14.437 & 0.050 & 14.487 \\
275 & $\mathrm{ScCo_{2}Fe_{3}}$ & 5.08645 & 4.97739 & 4.01446 & $+1.1999$ & 11.942 & 0.269 & 12.211 \\
276 & $\mathrm{ScCo_{5}}$ & 4.82248 & 4.82248 & 3.99718 & $+1.3039$ & 8.125 & 0.402 & 8.527 \\
277 & $\mathrm{SrNi_{2}Fe_{3}}$ & 5.37626 & 5.37626 & 4.13859 & $+1.8193$ & 10.533 & 0.141 & 10.674 \\
278 & $\mathrm{TbMn_{2}Fe_{3}}$ & 5.50829 & 5.50736 & 4.43387 & $-0.4031$ & 11.883 & -0.073 & 11.810 \\
279 & $\mathrm{TmGe_{2}Mn_{3}}$ & 5.85058 & 5.84954 & 4.13697 & $0.0000$ & 10.513 & -0.730 & 9.783 \\
280 & $\mathrm{TmMn_{2}Co_{3}}$ & 5.42699 & 5.42309 & 4.02095 & $0.0000$ & 11.960 & -0.380 & 11.580 \\
281 & $\mathrm{TmMn_{2}Fe_{3}}$ & 5.47311 & 5.47021 & 4.52846 & $0.0000$ & 16.327 & -0.706 & 15.621 \\
282 & $\mathrm{TmNi_{2}Fe_{3}}$ & 5.10720 & 5.10720 & 4.19253 & $0.0000$ & 8.730 & -0.464 & 8.266 \\
283 & $\mathrm{YCo_{2}Fe_{3}}$ & 5.28679 & 5.28668 & 4.11602 & $0.0000$ & 13.062 & 0.251 & 13.313 \\
284 & $\mathrm{YFe_{2}Co_{3}}$ & 5.52636 & 5.52263 & 4.06557 & $0.0000$ & 11.467 & 0.277 & 11.744 \\
285 & $\mathrm{YNi_{2}Fe_{3}}$ & 5.33964 & 5.33909 & 4.02144 & $+2.3033$ & 10.468 & 0.148 & 10.616 \\
286 & $\mathrm{YTc_{2}Fe_{3}}$ & 5.48394 & 5.48477 & 4.38209 & $-2.0092$ & 15.045 & 0.147 & 15.192 \\
\end{longtable}\endgroup

\bibliographystyle{unsrt}
\bibliography{references}